\documentclass[smallextended]{svjour3}     % onecolumn (second format) (Shifflett chose this one)
\begin{document}
%\large
\newcommand{\fl}{}
\newcommand{\ExpandDerivations}{0}

\title{A modification of Einstein-Schr\"{o}dinger theory that
       contains Einstein-Maxwell-Yang-Mills theory
%\thanks{Grants or other notes
%about the article that should go on the front page should be
%placed here. General acknowledgments should be placed at the end of the article.}
}
%\subtitle{Do you have a subtitle?\\ If so, write it here}

\titlerunning{A modification of Einstein-Schr\"{o}dinger theory}  % if too long for running head

\author{J. A. Shifflett}

\authorrunning{J. A. Shifflett} % if too long for running head

\institute{J. A. Shifflett\\
           Department of Physics, Washington University, \\
           1 Brookings Drive, St.~Louis, Missouri 63130, USA\\
\email{jashiffl@wustl.edu}
}

\date{Nov 9, 2008} %\today or an explicit date
%\date{Submitted: 31 July 2008, \ Accepted: 2 January 2009}
%\date{\today}
% The correct dates will be entered by the editor

\maketitle

\begin{abstract}
The Lambda-renormalized Einstein-Schr\"{o}dinger theory is a
modification of the original Einstein-Schr\"{o}dinger theory in which
a cosmological constant term is added to the Lagrangian, and it
has been shown to closely approximate Einstein-Maxwell theory.
Here we generalize this theory to non-Abelian fields by
letting the fields be composed of $d\times d$ Hermitian matrices.
The resulting theory incorporates the $U\!(1)$ and $SU\!(d)$ gauge terms of
Einstein-Maxwell-Yang-Mills theory, and is invariant under $U\!(1)$ and $SU\!(d)$ gauge transformations.
The special case where symmetric fields are multiples of the identity matrix
closely approximates Einstein-Maxwell-Yang-Mills theory
in that the extra terms in the field equations
are $<\!10^{-13}$ of the usual terms for worst-case fields accessible to measurement.
The theory contains a symmetric metric and Hermitian vector potential,
and is easily coupled to the additional fields of
Weinberg-Salam theory or flipped SU(5) GUT theory.
We also consider the case where symmetric fields have small traceless parts,
and show how this suggests a possible dark matter candidate.
\end{abstract}

\newcommand{\allowdisplaybreaks}{}
\newcommand{\Ax}{Appendix~}
\newcommand{\rmt}{\sqrt{2}\,i}
\newcommand{\ca}{c_1}
\newcommand{\cb}{c_2}
\newcommand{\cc}{c_3}
\newcommand{\sR}{{{^*}R}}
\newcommand{\dN}{d}
\newcommand{\W}{W}
\newcommand{\Wb}{\mathbf{W}}
\newcommand{\gb}{\mathbf{g}}
\newcommand{\gmat}{\gamma}
\newcommand{\hR}{\mathcal R}
\newcommand{\hhR}{\hat{\mathcal R}}
\newcommand{\tR}{\tilde\hR}
\newcommand{\hG}{\mathcal G}
\newcommand{\tG}{\tilde G}
\newcommand{\tB}{R}
\newcommand{\tC}{\acute C}
\newcommand{\tT}{\tilde T}
\newcommand{\tS}{\tilde S}
\newcommand{\cUps}{\check\Upsilon}
\newcommand{\bUps}{\bar\Upsilon}
\newcommand{\tGam}{\tilde\Gamma}
\newcommand{\cGam}{\check\Gamma}
\newcommand{\nGam}{\widehat\Gamma}
\newcommand{\sGam}{{^*\Gamma}}
\newcommand{\ftilde}{\tilde f}
\newcommand{\dual}{\vartheta}
\newcommand{\tdual}{\tilde\vartheta}
\newcommand{\Fdash}{\raise1pt\hbox{\rlap\textendash} F}
\newcommand{\Idash}{\raise1pt\hbox{\rlap\textendash} I}
\newcommand{\uacute}{\acute{u}}
\newcommand{\hz}{\bar h}
\newcommand{\hf}{\hat f}
\newcommand{\hj}{\hat j}
\newcommand{\hjj}{\hat{\mathbf j}}
\newcommand{\NPDEL}{\Delta}
\newcommand{\NPdel}{\delta}
\newcommand{\NPdels}{\delta^*}
\newcommand{\hQ}{\hat Q}
\newcommand{\J}{\hat j}
\newcommand{\sg}{\textsf{g}}
\newcommand{\g}{\emph{g}}
\newcommand{\bg}{\bar g}
\newcommand{\rmbg}{\sqrt{-\smash{\bar g}}}
\newcommand{\rmsg}{\sqrt{-\textsf{g}}}
\newcommand{\rmg}{\sqrt{-\emph{g}}}
\newcommand{\rmet}{\sqrt{-\eta}}
\newcommand{\rmN}{\sqrt{\!-N}}
\newcommand{\RmN}{\bar N}
\newcommand{\Rmg}{\bar\textsf{g}}
\newcommand{\rmgt}{\sqrt{-g_\diamond}}
\newcommand{\rmNt}{\sqrt{-N_\diamond}}
\newcommand{\dete}{{\bf e}}
\newcommand{\ehat}{Q}
\newcommand{\Isigma}{I}
\newcommand{\ff}{\,\ell}
\newcommand{\fog}{\hat r}
\newcommand{\Aphi}{A}
\newcommand{\Am}{{\mathcal A}}
\newcommand{\OPMU}{(1\!\pm\!\tanht)}
\newcommand{\OMPU}{(1\!\mp\!\tanht)}
\newcommand{\OPMIV}{(1\!\pm\!i\tant)}
\newcommand{\OMPIV}{(1\!\mp\!i\tant)}
\newcommand{\OPU}{(1\!+\!\tanht)}
\newcommand{\OMU}{(1\!-\!\tanht)}
\newcommand{\OPIV}{(1\!+\!i\tant)}
\newcommand{\OMIV}{(1\!-\!i\tant)}
\newcommand{\Gk}{G}
\newcommand{\FO}{O}
\newcommand{\NPD}{D}
\newcommand{\HD}{\varpi}
\newcommand{\Zht}{\check z}
\newcommand{\sinht}{\check{s}}
\newcommand{\cosht}{\check{c}}
\newcommand{\tanht}{\check{u}}
\newcommand{\cost}{\grave c}
\newcommand{\sint}{\grave s}
\newcommand{\tant}{\grave u}
\newcommand{\Zt}{\grave z}
\newcommand{\extra}{V}
\newcommand{\Fdashoverf}{\raise1pt\hbox{\rlap\textendash} I}
\newcommand{\btgam}{\lower2pt\hbox{\rlap\textendash} \tilde\gamma}
\newcommand{\mum}{\mu}
\newcommand{\Ddag}{\overleftarrow D}
\newcommand{\ord}{{\mathcal O}}
\newcommand{\sixteenpid}{2}
\newcommand{\eightpi}{2}
\newcommand{\fourpi}{ }
\newcommand{\nofourpi}{4\pi}
\newcommand{\HR}{\Re}

\def\Stacksymbols #1#2#3#4{\def\theguybelow{#2}
   \def\verticalposition{\lower#3pt}
   \def\spacingwithinsymbol{\baselineskip0pt\lineskip#4pt}
   \mathrel{\mathpalette\intermediary#1}}
\def\intermediary#1#2{\verticalposition\vbox{\spacingwithinsymbol
     \everycr={}\tabskip0pt
     \halign{$\mathsurround0pt#1\hfil##\hfil$\crcr#2\crcr
            \theguybelow\crcr}}}

% Einstein Maxwell spacetimes (04.40.Nr)
% cosmological constant (98.80.Es)
% dark matter (95.35.+d)
% exact solutions of general relativity (04.20.Jb)
% unified field theories and models (12.10.-g)
% relativity and gravitation (95.30.Sf) - probably covered by (04.40.Nr)
% alternative theories of gravity (04.50.+h)
% No more than 4 of these are supposed to be used
%\pacs{04.40.Nr,98.80.Es,12.10.-g,04.50.+h}% PACS codes
%\keywords{Einstein-Schrodinger Theory,
%Hermitian Theory of Relativity,
%Schrodinger Affine Field Theory,
%Einstein-Straus Theory,
%Einstein-Maxwell-Yang-Mills Theory,
%Cosmological Constant,
%Zero-Point Fluctuations
%}
%Use showkeys class option

%\ead{shifflet@hbar.wustl.edu}
\pagenumbering{arabic}

\section{\label{Introduction}Introduction}

The Einstein-Schr\"{o}dinger theory is a generalization of
vacuum general relativity which allows non-symmetric fields.
The theory without a cosmological constant was first proposed by Einstein and
Straus\cite{EinsteinStraus,Einstein3,EinsteinBianchi,EinsteinKaufman,EinsteinMOR}.
Schr\"{o}dinger later showed that the theory could be derived from a very simple Lagrangian
density\cite{SchrodingerI,SchrodingerIII,SchrodingerSTS}
if a cosmological constant was included.
Einstein and Schr\"{o}dinger suspected that the theory might include electrodynamics,
but no Lorentz force was found\cite{Callaway,Infeld}
when using the Einstein-Infeld-Hoffmann (EIH) method\cite{EinsteinInfeld,Wallace}.

In a previous paper\cite{xShifflett} we presented a simple modification of the
Einstein-Schr\"{o}dinger theory that contains Einstein-Maxwell theory.
The Lorentz force definitely results from the EIH method,
and in fact the ordinary Lorentz force equation results when sources are included.
The field equations match the ordinary Einstein and Maxwell equations except
for extra terms which are $<\!10^{-16}$ of the usual terms for
worst-case field strengths and rates-of-change accessible to measurement.
An exact electric monopole solution matches the
Reissner-Nordstr\"{o}m solution except for additional terms which are $<10^{-65}$ of
the usual terms for worst-case radii accessible to measurement.
An exact electromagnetic plane-wave solution
is identical to its counterpart in Einstein-Maxwell theory.
The modification of the original Einstein-Schr\"{o}dinger theory is the addition of a
second cosmological term $\Lambda_z g_{\mu\nu}$, where $g_{\mu\nu}$ is the symmetric metric.
We assume this term is nearly canceled by
Schr\"{o}dinger's ``bare'' cosmological term $\Lambda_b N_{\mu\nu}$, where $N_{\mu\nu}$ is the
nonsymmetric fundamental tensor. The total ``physical'' cosmological constant
$\Lambda =\Lambda_b+\Lambda_z$ can then match measurements of the accelerating universe.
A possible origin of our $\Lambda_z$ is from zero-point fluctuations\cite{Zeldovich,Sahni,Peskin,Carroll}
and Higgs field vacuum energy, although we just take $\Lambda_z$ as given, without regard to its origin.
The theory in \cite{xShifflett} is related to one in \cite{Kursunoglu},
but it is roughly the electromagnetic dual of that theory,
it uses a different nonsymmetric Ricci tensor with special transformation properties,
it allows coupling to additional fields (sources), and it allows $\Lambda\ne0$.

Here we generalize the theory in \cite{xShifflett} to non-Abelian fields
by letting the fields be composed of $d\times d$ Hermitian matrices\cite{ShifflettThesis}.
This is done much as it is done in \cite{Borchsenius76,Borchsenius78} with Bonnor's theory\cite{Bonnor}.
The resulting theory incorporates the $U\!(1)$ and $SU\!(d)$ gauge terms
of the Einstein-Maxwell-Yang-Mills Lagrangian,
and if we assume that symmetric fields are multiples of the identity matrix,
we get a close approximation to Einstein-Maxwell-Yang-Mills theory.
The theory can be coupled to
additional fields using a symmetric metric $g_{\mu\nu}$ and Hermitian vector potential $\Am_\mu$.
If we let $d\!=\!2$ and couple the theory to the Standard Model,
the $U\!(1)$ and $SU\!(2)$ gauge terms are incorporated together with
the geometry, and the combined theory is invariant under
$U\!(1)\otimes SU\!(2)\otimes SU\!(3)$.
Likewise, if we let $d\!=\!5$ and couple the theory to flipped $SU\!(5)$ GUT theory,
the $U\!(1)$ and $SU\!(5)$ gauge terms are incorporated together with
the geometry, and the theory is invariant under $U\!(1)\otimes SU\!(5)$.
Note that flipped $SU\!(5)$ GUT theory\cite{Barr,Ellis} avoids the short
proton lifetime and other problems of the original $SU\!(5)$ GUT theory.
Assuming that we use the usual fermion and Higgs field Lagrangian
we will get the usual energy-momentum tensor in the Einstein equations,
the usual current in Ampere's law,
the usual equations of motion for fermion and Higgs fields,
and the usual mixing and mass acquisition in $\Am_\mu$.

This paper is organized as follows. In \S\ref{LagrangianDensity} we discuss the Lagrangian density.
In \S\ref{InvarianceProperties} we show that the Lagrangian density is real and invariant under
$U\!(1)$ and $SU\!(d)$ gauge transformations.
In \S\ref{NonsymmetricCase} we consider the special case where the symmetric
fields are multiples of the identity matrix,
and we quantify how closely this theory approximates Einstein-Maxwell-Yang-Mills theory.
In \S\ref{SymmetricCase} we consider the case where the symmetric fields
have small traceless components, and show how this suggests a possible dark matter candidate.

\section{\label{LagrangianDensity}The Lagrangian density}
Einstein-Maxwell-Yang-Mills theory can be derived from a Palatini Lagrangian,
\begin{eqnarray}
\label{GR}
\fl {\mathcal L}(\Gamma^{\lambda}_{\rho\tau},g_{\rho\tau},\Am_\nu)
&=&-\frac{\lower2pt\hbox{$1$}}{16\pi}\rmg\left[\,g^{\mu\nu}
R_{\nu\mu}({\Gamma})+(n\!-\!2)\Lambda_b\,\right]\nonumber\\
&&\,+\frac{\lower2pt\hbox{$1$}}{8\pi}\rmg\,\, \frac{tr(F_{\rho\alpha}g^{\alpha\mu}g^{\rho\nu}\!F_{\nu\mu})}{2d}
+{\mathcal L}_m(g_{\mu\nu},\Am_\nu,\psi,\phi\cdots),~~~~
\end{eqnarray}
containing a metric $g_{\nu\mu}$, connection $\Gamma^\alpha_{\!\nu\mu}$,
and Maxwell-Yang-Mills field tensor
\begin{eqnarray}
\label{FYangMills}
F_{\nu\mu}\!&=&\!2\Am_{[\mu,\nu]}+i[\Am_\nu,\Am_\mu]g_c/\hbar\sqrt{2d}.
\end{eqnarray}
Here $\Lambda_b$ is a cosmological constant and $g_c$ is the coupling constant.
The vector potential $\Am_\sigma$ is composed of
$d\!\times\!d$ Hermitian matrices and can be decomposed into a real $U\!(1)$
gauge vector $A^0_\sigma$ and $d^{\,2}\!-\!1$ real $SU\!(d)$ gauge vectors $A^a_\nu$,
\begin{eqnarray}
\label{Amsplit}
\fl \Am_\nu&=&\Isigma\Aphi^0_\nu+\tau_a A^a_\nu.
\end{eqnarray}
Here $I$ is the identity matrix and the generators $\tau_a$ are $d\!\times\!d$ matrices with
\begin{eqnarray}
\label{tauproperties}
\fl &&[\tau_a,\tau_b]=i\sqrt{2d}\,f_{abc}\tau_c,
~~\tau_a^*=\tau_a^T,
~~tr(\tau_a)=0,
~~tr(\tau_a\tau_b)=d\delta^a_b,~~~~~
\end{eqnarray}
where the $f_{abc}$ are totally antisymmetric structure constants.
For example, with $d\!=\!2$ the $\tau_a$ are the Pauli matrices,
$f_{abc}\!=\!\epsilon_{abc}$, and
$g_c\!=\!e/sin\theta_w$ where $\theta_w$ is the weak mixing angle.
The ${\mathcal L}_m$ term couples $g_{\mu\nu}$ and $\Am_\mu$ to additional fields,
as in Weinberg-Salam theory
%(see Appendix \ref{WeinbergSalam})
or flipped $SU\!(5)$ GUT theory.
%(see Appendix \ref{FlippedSU5}).
The symbols $\raise2pt\hbox{$_{(~)}$}$ and $\raise2pt\hbox{$_{[~]}$}$ around indices
indicate symmetrization and antisymmetrization, and $[A,B]\!=\!AB\!-\!BA$.
Note that the term $tr(F_{\rho\alpha}g^{\alpha\mu}g^{\rho\nu}\!F_{\nu\mu})$
in (\ref{GR}) contains both $U\!(1)$ and $SU\!(d)$ gauge terms.
Our conventions differ a bit from the usual ones.
The factors of $2d$ in (\ref{GR},\ref{FYangMills}) result because
our $\tau_a$ are normalized like the identity matrix
as in (\ref{tauproperties}) instead of with $tr(\tau_a\tau_b)\!=\!\delta^a_b/2$,
so that $\Aphi^0$ and $A^a$ are on an equal footing in (\ref{Amsplit}).
As in \cite{xShifflett}, a factor of $1/4\pi$ in (\ref{GR}) results because
we are not using the Heaviside-Lorentz convention.
%where electromagnetic fields are scaled down by $1/\sqrt{4\pi}$
%and coupling constants are scaled up by $\sqrt{4\pi}$.
%To put $\Aphi^0$ and $A^a$ on an equal footing in (\ref{Amsplit}),
%the $\tau_a$ are normalized like the identity matrix
%as in (\ref{tauproperties}) instead of with $tr(\tau_a\tau_b)\!=\!1/2$, and
%this causes the $1/2d$ in (\ref{GR},\ref{FYangMills}).
%We are not using the
%Heaviside-Lorentz convention, which causes a factor of $1/4\pi$ in (\ref{GR}).
We are also using geometrized units with $c\!=\!G\!=\nobreak\!1$ instead of
natural units with $c\!=\!\hbar\!=\nobreak\!1$.

The original Einstein-Schr\"{o}dinger theory
allows a nonsymmetric $N_{\mu\nu}$ and $\nGam^{\lambda}_{\!\rho\tau}$
in place of the symmetric $g_{\mu\nu}$ and $\Gamma^{\lambda}_{\rho\tau}$, and
excludes the $tr(F_{\rho\alpha}g^{\alpha\mu}g^{\rho\nu}\!F_{\nu\mu})$ term.
Our theory introduces an additional cosmological term $\Rmg  \Lambda_z$ as in \cite{xShifflett},
and also allows $\nGam^\rho_{\nu\mu}$ and $N_{\nu\mu}$ to have
$d\!\times\!d$ matrix components\cite{ShifflettThesis},
\begin{eqnarray}
\label{JSlag4}
\fl {\mathcal L}(\nGam^\alpha_{\rho\tau},N_{\rho\tau})
&=&-\frac{\lower2pt\hbox{$1$}}{16\pi d}\,\RmN [\,tr(N^{\dashv\mu\nu}\hhR_{\nu\mu})
+d(n\!-\!2)\Lambda_b\,]\nonumber\\
\fl &&-\frac{\lower2pt\hbox{$1$}}{16\pi}\,\Rmg (n\!-\!2)\Lambda_z
+{\mathcal L}_m(\sg_{\mu\nu},\Am_\nu,\psi,\phi\cdots),
\end{eqnarray}
where $\Lambda_b\!\approx\!-\Lambda_z$ so that the total $\Lambda$ matches astronomical measurements\cite{Astier}
\begin{eqnarray}
\label{Lambdadef}
\Lambda\!=\!\Lambda_b\!+\!\Lambda_z\!\approx\!10^{-56} cm^{-2},
\end{eqnarray}
and the vector potential is defined to be
\begin{eqnarray}
\label{A2}
\Am_\nu={\nGam}^\sigma_{\![\nu\sigma]}/[(n-\!1)\,i\sqrt{\sixteenpid\Lambda_b}\,].
\end{eqnarray}
The ``physical'' metric $\g^{\nu\mu}$ and the fields $\sg^{\nu\mu}$, $\hz^{\nu\mu}$, $\Rmg$ and $\RmN$ are defined by
\begin{eqnarray}
\label{sgdef2}
\fl\Rmg\sg^{\nu\mu}=\RmN\!N^{\dashv(\nu\mu)},~~
~~~\Rmg\sg^{\nu\mu}\!=\!\rmg\,(Ig^{\nu\mu}\!-\! \hz^{\nu\mu}),
~~~tr(\hz^{\nu\mu})=0,~~~~~~~~~~~~~~~~\\
\label{RmgRmNdef}
 \Rmg\!=\!(\pm det(\sg_{\nu\mu}))^{1/2d},~~\RmN\!=\!(\pm det(N_{\nu\mu}))^{1/2d},~~
  {\rm +~for~even~d,-~for~odd~d.}~~~
\end{eqnarray}
Note that (\ref{sgdef2}) defines $\sg^{\mu\nu}$ unambiguously
because $\Rmg\!=\![\pm det(\Rmg\sg^{\mu\nu})]^{1/d(n-2)}$.
The symmetric metric $g^{\nu\mu}$ is used for measuring space-time intervals, covariant derivatives,
and for raising and lowering indices.
The ${\mathcal L}_m$ term is not to include a $tr(F_{\rho\alpha}g^{\alpha\mu}g^{\rho\nu}\!F_{\nu\mu})$ term
but may contain source terms with the usual coupling to $\Am_\nu$ and $g_{\nu\mu}$.
Tensor indices are assumed to have dimension n=4, but as with the matrix dimension ``d'',
we will retain ``n'' in the equations to show how easily the theory can be generalized.
The non-Abelian Ricci tensor in (\ref{JSlag4})
is chosen to have special symmetry properties to be discussed later,
\begin{eqnarray}
\label{HermitianizedRicci2}
\fl \hhR_{\nu\mu}=\nGam^\alpha_{\nu\mu,\alpha}
\!-\nGam^\alpha_{\!(\alpha(\nu),\mu)}
\!+\frac{\lower2pt\hbox{$1$}}{2}\nGam^\sigma_{\nu\mu}\nGam^\alpha_{\!(\sigma\alpha)}
\!+\frac{\lower2pt\hbox{$1$}}{2}\nGam^\alpha_{\!(\sigma\alpha)}\nGam^\sigma_{\nu\mu}
\!-\nGam^\sigma_{\nu\alpha}\nGam^\alpha_{\sigma\mu}
\!-\frac{{\nGam}{^\tau_{\![\tau\nu]}}{\nGam}{^\rho_{\![\rho\mu]}}}{(n\!-\!1)}\,.~~~~
\end{eqnarray}
For Abelian fields the third and fourth terms are the same,
and this tensor reduces to the Abelian version in \cite{xShifflett}.
This tensor reduces to the ordinary Ricci tensor for $\nGam^\alpha_{[\nu\mu]}\!=\!0$
and $\nGam^\alpha_{\alpha[\nu,\mu]}\!=\!0$, as occurs in ordinary general relativity.

The determinants $\sg\!=\!det(\sg_{\nu\mu})$ and
$N\!=\!det(N_{\nu\mu})$ are defined as usual but where $N_{\nu\mu}$ and $\sg_{\nu\mu}$ are taken to
be $nd\times nd$ matrices.
The inverse of $N_{\nu\mu}$ is defined to be
$N^{\dashv\mu k\nu i}\!=\!(1/N)\partial N/\partial N_{\nu i\mu k}$
where i,k are matrix indices,
or $N^{\dashv\mu\nu}\!=\!(1/N)\partial N/\partial N_{\nu\mu}$ using matrix notation.
The field $N^{\dashv\mu\nu}$ satisfies the relation
$N^{\dashv\mu k\nu i}\!N_{\nu i\sigma j}\!=\nobreak\!\delta^\mu_\sigma\delta^k_j$,
or $N^{\dashv\mu\nu}\!N_{\nu\sigma}\!=\nobreak\!\delta^\mu_\sigma I$ using matrix notation.
Likewise $\sg_{\nu\sigma}$ is the inverse of $\sg^{\mu\nu}$ such that
$\sg^{\mu\nu}\!\sg_{\nu\sigma}\!=\nobreak\!\delta^\mu_\sigma I$.
Assuming $\tilde N_{\alpha\tau}\!=\!T^\nu_\alpha N_{\nu\mu}T^\mu_\tau$
for some coordinate transformation $T^\nu_\alpha\!=\!\partial x^\nu/\partial \tilde x^\alpha$,
the transformed determinant $\tilde N\!=\!det(\tilde N_{\alpha\tau})$ will contain $d$ times as many
$T^\nu_\alpha$ factors as it would if $N_{\alpha\tau}$ had no matrix components,
so $N$ and $\sg$ are scalar densities of weight 2d.
The factors $\RmN$ and $\Rmg$ from (\ref{RmgRmNdef}) are used in (\ref{JSlag4})
instead of $\rmN$ and $\rmg$ to make the Lagrangian density a scalar density
of weight~1 as required.

For our theory the Maxwell-Yang-Mills field tensor $f^{\nu\mu}$ is defined by
\begin{eqnarray}
\label{fdef2}
\Rmg f^{\nu\mu}=i\RmN\!N^{\dashv[\nu\mu]}\Lambda_b^{\!1/2}/\sqrt{\sixteenpid}.
\end{eqnarray}
Then from (\ref{sgdef2}), $\sg^{\mu\nu}$ and $f^{\mu\nu}$ are parts of a total field,
\begin{eqnarray}
\label{Wdef2}
(\RmN/\Rmg)N^{\dashv\nu\mu}=\sg^{\mu\nu}\!+\!if^{\mu\nu}\sqrt{\sixteenpid}\,\Lambda_b^{\!-1/2}.
\end{eqnarray}
We will see that the field equations require
$f_{\nu\mu}\!\approx\!2\Am_{[\mu,\nu]}+i\sqrt{\sixteenpid\Lambda_b}\,[\Am_\nu,\Am_\mu]$ to a very high precision.
From (\ref{FYangMills}) this agrees with Einstein-Maxwell-Yang-Mills theory when
$\sqrt{\sixteenpid\Lambda_b}\!=\!g_c/\hbar\sqrt{2d}$. Using $d\!=\!2$, $g_c\!=\!e/sin\theta_w$ and (\ref{Lambdadef}) gives
\begin{eqnarray}
\label{Lambdab}
-\Lambda_z\approx\Lambda_b\!=\!\frac{1}{4d}\!\left(\frac{g_c}{\hbar}\right)^2
=\frac{\lower1pt\hbox{$\alpha$}}{4dl_P^2}\left(\frac{g_c}{e}\right)^2
=1.5\times 10^{63}cm^{-2},~~~
\end{eqnarray}
where $l_P\!=\!\sqrt{G\hbar/c^3}\!=\!1.616\!\times\!10^{-33}cm$,
$\alpha\!=\!e^2/\fourpi\hbar c\!=\!1/137$ and
$\sin^2\theta_w\!=\!.23$.

It is helpful to decompose $\nGam^\rho_{\!\nu\mu}$ into a new connection $\tGam{^\alpha_{\!\nu\mu}}$,
and $\Am_\nu$ from (\ref{A2}),
\begin{eqnarray}
\label{gamma_natural2}
\fl {\nGam}{^\alpha_{\nu\mu}}&=&\tGam{^\alpha_{\nu\mu}}
+(\delta^\alpha_\mu\Am_\nu
\!-\delta^\alpha_\nu \Am_\mu)\,i\sqrt{\sixteenpid\Lambda_b}\,,\\
\label{gamma_tilde2}
\fl {\rm where}~~~\tGam{^\alpha_{\nu\mu}}
&=&{\nGam}{^\alpha_{\nu\mu}}\!+
(\delta^\alpha_\mu{\nGam}{^\sigma_{\![\sigma\nu]}}
-\delta^\alpha_\nu{\nGam}{^\sigma_{\![\sigma\mu]}})/(n\!-\!1).
\end{eqnarray}
By contracting (\ref{gamma_tilde2}) on the right and left we see that
$\tGam{^\alpha_{\nu\mu}}$ has the symmetry
\begin{eqnarray}
\label{JScontractionsymmetric2}
\tGam^\alpha_{\nu\alpha}
\!=\!\nGam^\alpha_{\!(\nu\alpha)}
\!=\!\tGam^\alpha_{\alpha\nu},
\end{eqnarray}
so it has only $n^3\!-\!n$ independent components whereas $\nGam{^\alpha_{\nu\mu}}$ has $n^3$.
Substituting the decomposition (\ref{gamma_natural2}) into (\ref{HermitianizedRicci2})
using (\ref{breakoutA}) from Appendix \ref{hRproperties},
\begin{eqnarray}
\label{breakout0}
\fl \hR_{\nu\mu}(\nGam)=\hR_{\nu\mu}(\tGam)
&+&2\Am_{[\nu,\mu]}\,i\sqrt{\sixteenpid\Lambda_b}+\sixteenpid\Lambda_b[\Am_\nu,\Am_\mu]\nonumber\\
\fl &+&([\Am_\alpha,\tGam{^\alpha_{\nu\mu}}]
-[\Am_{(\nu},\tGam^\alpha_{\mu)\alpha}])\,i\sqrt{\sixteenpid\Lambda_b}\,.~~~~~
\end{eqnarray}
Using (\ref{breakout0}), the Lagrangian density (\ref{JSlag4})
can be rewritten in terms of ${\tGam}{^\alpha_{\nu\mu}}$ and $\Am_\sigma$
from (\ref{gamma_tilde2},\ref{A2}),
\begin{eqnarray}
\label{JSlag5}
\fl {\mathcal L}
&=&\!-\frac{\lower1pt\hbox{$1$}}{16\pi d}\RmN\!\left[\,tr(N^{\dashv\mu\nu}(\tR_{\nu\mu}
\!+2\Am_{[\nu,\mu]}\,i\sqrt{\sixteenpid\Lambda_b}
\!+\sixteenpid\Lambda_b[\Am_\nu,\Am_\mu]\right.\nonumber\\
\fl &&~~~~~~~~~~~~~~~~~~~~~~+\left.([\Am_\alpha,\tGam{^\alpha_{\nu\mu}}]
-[\Am_{(\nu},\tGam^\alpha_{\mu)\alpha}])\,i\sqrt{\sixteenpid\Lambda_b}\,))
+d(n\!-\!2)\Lambda_b\right]\nonumber\\
\fl &&-\frac{\lower1pt\hbox{$1$}}{16\pi}\,\Rmg  (n\!-\!2)\Lambda_z
+{\mathcal L}_m(\sg_{\mu\nu},\Am_\sigma,\psi,\phi\dots).
\end{eqnarray}
Here $\tR_{\nu\mu}\!=\!\hR_{\nu\mu}(\tGam)$, and from (\ref{JScontractionsymmetric2})
our non-Abelian Ricci tensor (\ref{HermitianizedRicci2}) reduces~to
\begin{eqnarray}
\label{HermitianizedRiccit2}
\fl \tR_{\nu\mu}
&=&\tGam^\alpha_{\nu\mu,\alpha}
-\tGam^\alpha_{\alpha(\nu,\mu)}
+\frac{\lower2pt\hbox{$1$}}{2}\tGam^\sigma_{\nu\mu}\tGam^\alpha_{\sigma\alpha}
+\frac{\lower2pt\hbox{$1$}}{2}\tGam^\alpha_{\sigma\alpha}\tGam^\sigma_{\nu\mu}
-\tGam^\sigma_{\nu\alpha}\tGam^\alpha_{\sigma\mu}.
\end{eqnarray}
From (\ref{gamma_natural2},\ref{JScontractionsymmetric2}), $\tGam^\alpha_{\nu\mu}$ and $\Am_\nu$ fully
parameterize $\nGam^\alpha_{\nu\mu}$ and can be treated as independent variables.
The fields $\RmN\!N^{\dashv(\nu\mu)}$ and $\RmN\!N^{\dashv[\nu\mu]}$
(or~$\sg^{\nu\mu}$ and~$f^{\nu\mu}$)
fully parameterize $N_{\nu\mu}$ and can also be treated as independent variables.
It is simpler to calculate the field equations by setting
$\delta{\mathcal L}/\delta\tGam^\alpha_{\nu\mu}\!=\!0$,
$\delta{\mathcal L}/\delta \Am_\nu\!=\!0$,
$\delta{\mathcal L}/\delta(\RmN\!N^{\dashv(\mu\nu)})\!=\!0$ and
$\delta{\mathcal L}/\delta(\RmN\!N^{\dashv[\mu\nu]})\!=\nobreak\!\nobreak 0$
instead of setting $\delta{\mathcal L}/\delta\nGam^\alpha_{\nu\mu}\!=\nobreak\!\nobreak 0$ and
$\delta{\mathcal L}/\delta N_{\nu\mu}\!=\!0$,
so we will follow this method.

\section{\label{InvarianceProperties}Invariance properties of the Lagrangian density}
Here we show that the Lagrangian density is real (invariant under complex conjugation),
and is also invariant under $U\!(1)$ and $SU\!(d)$ gauge transformations.
The Abelian Lambda-renormalized Einstein-Schr\"{o}dinger theory comes in two versions,
one where $\nGam^\rho_{\nu\mu}$ and $N_{\nu\mu}$ are real, and one where they are Hermitian.
The non-Abelian theory also comes in two versions, one where $\nGam^\rho_{\nu\mu}$ and
$N_{\nu\mu}$ are real, and one where they have $nd\!\times\!nd$ Hermitian symmetry,
$\nGam^{\alpha\,*}_{\nu i\mu k}\!=\!\nGam^\alpha_{\mu k\nu i}$
and $N^*_{\nu i\mu k}\!=\!N_{\mu k\nu i}$, where $i,k$ are matrix indices.
Using matrix notation these symmetries become
\begin{eqnarray}
\label{Hermitian2}
\fl \nGam^{\alpha\,*}_{\nu\mu}\!=\!\nGam^{\alpha\,T}_{\mu\nu},
~~~~\tGam^{\alpha\,*}_{\nu\mu}\!=\!\tGam^{\alpha\,T}_{\mu\nu},
~~~~N^*_{\nu\mu}\!=\!N^T_{\mu\nu},
~~~~N^{\dashv\mu\nu\,*}\!=\!N^{\dashv\nu\mu\,T},
\end{eqnarray}
where ``T'' indicates matrix transpose (not transpose over tensor indices).
We will assume this Hermitian case
because it results from $\Lambda_z\!<\!0,~\Lambda_b\!>\!0$ as in (\ref{Lambdab}).
From (\ref{Hermitian2},\ref{sgdef2},\ref{fdef2},\ref{A2}) the physical fields
are all composed of $d\!\times\!d$ Hermitian matrices,
\begin{eqnarray}
\fl \sg^{\nu\mu\,*}\!=\!\sg^{\nu\mu\,T}\!,~\sg_{\nu\mu}^*\!=\!\sg_{\nu\mu}^T,~
f^{\nu\mu\,*}\!=\!f^{\nu\mu\,T}\!,~f_{\nu\mu}^*\!=\!f_{\nu\mu}^T,~
\nGam^{\alpha\,*}_{\!(\nu\mu)}\!=\!\nGam^{\alpha\,T}_{\!(\nu\mu)},~
\Am_\nu^*\!=\!\Am_\nu^T.~~~
\end{eqnarray}
Hermitian $f_{\nu\mu}$ and $\Am_\nu$ are just what we need to approximate Einstein-Maxwell-Yang-Mills theory.
And of course $\sg^{\nu\mu}$ and $\sg_{\nu\mu}$ will be Hermitian if we assume the special case where
they are multiples of the identity matrix.
Writing the symmetries as $N^*_{\nu i\mu k}\!=\!N_{\mu k\nu i}$,
$~\sg^*_{\nu i\mu k}\!=\!\sg_{\nu k\mu i}\!=\!\sg_{\mu k\nu i}$,
and using the result that the determinant of a Hermitian matrix is real,
we see that the $nd\!\times nd$ matrix determinants are real
\begin{eqnarray}
\label{Ngreal}
N^*=N,~~~\sg^*=\sg,~~~g^*=g.
\end{eqnarray}
Also, using (\ref{Hermitian2}) and the identity $M_1^TM_2^T=(M_2M_1)^T$ we can deduce
a remarkable property of our non-Abelian Ricci tensor (\ref{HermitianizedRicci2}),
which is that it has the same $nd\!\times\!nd$ Hermitian symmetry as $\nGam^\alpha_{\nu\mu}$ and $N_{\nu\mu}$,
\begin{eqnarray}
\label{transpositionsymmetric2}
\hhR^*_{\nu\mu}=\hhR^T_{\mu\nu}.
\end{eqnarray}
From the properties (\ref{transpositionsymmetric2},\ref{Hermitian2},\ref{Ngreal})
and the identities $tr(M_1M_2)\!=\!tr(M_2M_1)$, $tr(M^T)\!=\!tr(M)$
we see that our Lagrangian density (\ref{JSlag4}) or (\ref{JSlag5}) is real.

With an $SU\!(d)$ gauge transformation we assume a transformation matrix $U$ that is
special $(det(U)\!=\!1)$ and unitary $(U^\dagger U\!=\!I)$.
Taking into account (\ref{Amsplit},\ref{A2},\ref{gamma_natural2}),
we assume that under an $SU\!(d)$ gauge transformation the fields transform as follows,
\begin{eqnarray}
\label{Btransform}
\fl \tau_a A^a_\nu&\rightarrow& U\tau_a A^a_\nu U^{-\!1} +\frac{\lower2pt\hbox{$i$}}{\sqrt{\sixteenpid\Lambda_b}}\,U_{,\nu}U^{-\!1},\\
\label{Amtransform}
\fl \Am_\nu&\rightarrow& U\Am_\nu U^{-\!1} +\frac{\lower2pt\hbox{$i$}}{\sqrt{\sixteenpid\Lambda_b}}\,U_{,\nu}U^{-\!1},\\
\label{Atransform}
\fl A^0_\nu&\rightarrow& A^0_\nu,\\
\label{gammatransform}
\fl {\nGam}{^\alpha_{\nu\mu}}&\rightarrow& U{\nGam}{^\alpha_{\nu\mu}}U^{-\!1}
+2\delta^\alpha_{[\nu}U_{,\mu]}U^{-\!1},\\
\fl {\nGam}{^\alpha_{(\nu\mu)}}&\rightarrow& U{\nGam}{^\alpha_{(\nu\mu)}}U^{-\!1},\\
\fl {\nGam}{^\alpha_{[\alpha\mu]}}&\rightarrow& U{\nGam}{^\alpha_{[\alpha\mu]}}U^{-\!1}
+(n\!-\!1)\,U_{,\mu}U^{-\!1},\\
\label{tGamtransform}
\fl \tGam^\alpha_{\nu\mu}&\rightarrow& U\tGam^\alpha_{\nu\mu}U^{-\!1},\\
\label{Ntransform}
\fl N_{\nu\mu}&\rightarrow& UN_{\nu\mu}U^{-\!1},~~~~~
\sg_{\nu\mu}\rightarrow U\sg_{\nu\mu}U^{-\!1},~~~~
f_{\nu\mu}\rightarrow Uf_{\nu\mu}U^{-\!1},\\
\label{NItransform}
\fl N^{\dashv\mu\nu}&\rightarrow& UN^{\dashv\mu\nu}U^{-\!1},~~~
\sg^{\mu\nu}\rightarrow U\sg^{\mu\nu}U^{-\!1},~~~~
f^{\mu\nu}\rightarrow Uf^{\mu\nu}U^{-\!1}.
\end{eqnarray}
Under a $U\!(1)$ gauge transformation all of the fields are unchanged except
\begin{eqnarray}
\label{AU1}
\fl A^0_\nu&\rightarrow& A^0_\nu +\frac{\lower2pt\hbox{$1$}}{\sqrt{\eightpi\Lambda_b}}\,\varphi_{,\nu},\\
\label{AmU1}
\fl \Am_\nu&\rightarrow& \Am_\nu +\frac{\lower2pt\hbox{$\Isigma$}}{\sqrt{\sixteenpid\Lambda_b}}\,\varphi_{,\nu},\\
\label{gammaU1}
\fl {\nGam}{^\alpha_{\nu\mu}}&\rightarrow& {\nGam}{^\alpha_{\nu\mu}}
-2i\Isigma\,\delta^\alpha_{[\nu}\varphi_{,\mu]},\\
\label{gammaantiU1}
\fl {\nGam}{^\alpha_{[\alpha\mu]}}&\rightarrow& {\nGam}{^\alpha_{[\alpha\mu]}}
-(n\!-\!1)i\Isigma\,\varphi_{,\mu}.
\end{eqnarray}

Writing the $SU\!(d)$ gauge transformation (\ref{Ntransform}) as
\begin{eqnarray}
\fl N'_{\nu\mu}
=\begin{pmatrix}{U&0&0&0\cr 0&U&0&0\cr 0&0&U&0\cr 0&0&0&U}\end{pmatrix}
\!\!\begin{pmatrix}{
N_{00}\!&\!N_{01}\!&\!N_{02}\!&\!N_{03}\cr
N_{10}\!&\!N_{11}\!&\!N_{12}\!&\!N_{13}\cr
N_{20}\!&\!N_{21}\!&\!N_{22}\!&\!N_{23}\cr
N_{30}\!&\!N_{31}\!&\!N_{32}\!&\!N_{33}}\end{pmatrix}
\!\!\begin{pmatrix}{
U^{-\!1}\!\!&0&0&0\cr
0&\!\!U^{-\!1}\!\!&0&0\cr
0&0&\!\!U^{-\!1}\!\!&0\cr
0&0&0&\!\!U^{-\!1}\!\!}\end{pmatrix}~~
\end{eqnarray}
and using the identity $det(M_1M_2)\!=\!det(M_1)det(M_2)$, we see that the
$nd\!\times nd$ matrix determinants are invariant under an $SU\!(d)$ gauge transformation,
\begin{eqnarray}
\label{determinanttransform}
N\rightarrow N,~~~\sg\rightarrow \sg,~~~g\rightarrow g.
\end{eqnarray}
Another remarkable property of our non-Abelian Ricci tensor (\ref{HermitianizedRicci2})
is that it transforms like $N_{\nu\mu}$ under an $SU\!(d)$ gauge transformation (\ref{gammatransform}),
as in (\ref{hRtransformA}) of Appendix \ref{hRproperties}
\begin{eqnarray}
\label{hRtransform}
\fl \hR_{\nu\mu}(U{\nGam}{^\alpha_{\rho\tau}}U^{-\!1}
\!+\!2\delta^\alpha_{[\rho}U_{,\tau]}U^{-\!1})
=U\hR_{\nu\mu}({\nGam}{^\alpha_{\rho\tau}})U^{-\!1}
~~~{\rm for~any~matrix}~U(x^\sigma).~~~~
\end{eqnarray}
The results (\ref{determinanttransform},\ref{hRtransform}) actually apply for any invertible matrix $U$,
and do not require that $det(U)\!=\!1$ or $U^\dagger U\!=\!I$. Using the special case $U=Ie^{-i\varphi}$
in (\ref{hRtransform}) we see that our non-Abelian Ricci tensor (\ref{HermitianizedRicci2})
is also invariant under a $U\!(1)$ gauge transformation,
\begin{eqnarray}
\label{gaugesymmetric2}
\hR_{\nu\mu}(\nGam^\alpha_{\rho\tau}\!-2i\Isigma\,\delta^\alpha_{[\rho}\varphi_{,\tau]})=
\hR_{\nu\mu}(\nGam^\alpha_{\rho\tau})~~~~~{\rm for~any}~\varphi(x^\sigma).
\end{eqnarray}
From (\ref{hRtransform},\ref{Ntransform},\ref{determinanttransform},\ref{gaugesymmetric2})
and the identity $tr(M_1M_2)\!=\!tr(M_2M_1)$ we see that our Lagrangian density (\ref{JSlag4})
or (\ref{JSlag5}) is invariant under both $U\!(1)$ and $SU\!(d)$ gauge transformations,
thus satisfying an important requirement to approximate Einstein-Maxwell-Yang-Mills theory.

One of the motivations for this theory is that the $\Lambda_z\!=\!0$, ${\mathcal L}_m\!=\!0$
version can be derived from a purely affine Lagrangian density as well as
a Palatini Lagrangian density, the same as with the Abelian theory\cite{SchrodingerI}.
The purely affine Lagrangian density is
\begin{eqnarray}
\label{JSlagaffine2}
{\mathcal L}(\nGam^\alpha_{\rho\tau})=[\pm det(N_{\nu\mu})]^{1\!/2d},
\end{eqnarray}
where $N_{\nu\mu}$ is simply defined to be
\begin{eqnarray}
\label{Naffine2}
N_{\nu\mu}=-\hhR_{\nu\mu}/\Lambda_b.
\end{eqnarray}
Considering that $N^{\dashv\mu\nu}=(1/N)\partial N/\partial N_{\nu\mu}$,
we see that setting $\delta{\mathcal L}/\delta\nGam^\alpha_{\rho\tau}\!=\!0$ gives the
same result obtained from (\ref{JSlag4}) with $\Lambda_z\!=\!0$, ${\mathcal L}_m\!=\!0$,
\begin{eqnarray}
tr[N^{\dashv\mu\nu}\delta\hhR_{\nu\mu}/\delta\nGam^\alpha_{\rho\tau}]=0.
\end{eqnarray}
Since (\ref{JSlagaffine2}) depends only on $\nGam^\alpha_{\rho\tau}$, there are no
$\delta{\mathcal L}/\delta(\RmN\!N^{\dashv\mu\nu})\!=\!0$ field equations.
However, the definition (\ref{Naffine2}) exactly matches
the $\delta{\mathcal L}/\delta(\RmN\!N^{\dashv\mu\nu})\!=\!0$ field equations
obtained from (\ref{JSlag4}) with $\Lambda_z\!=\!0$, ${\mathcal L}_m\!=\!0$.

Note that there are other definitions of $N$ and $g$ which would make the Lagrangian
density (\ref{JSlag4}) real and gauge invariant,
for example we could have defined $N\!\nobreak=\nobreak\!tr(\rm{det}(N_{\nu\mu}))$
or $N\!\nobreak=\nobreak\!\rm{Det}(\rm{det}(N_{\nu\mu}))$,
where $\rm{det()}$ is done only over the tensor indices. However, with these definitions
the field $N^{\dashv\mu\nu}\!=\!(1/N)\partial N/\partial N_{\nu\mu}$ would not be a
matrix inverse such that $N^{\dashv\sigma\nu}\!N_{\nu\mu}\!=\nobreak\!\delta^\sigma_\mu I$.
Calculations would be very unwieldy in a theory where $N^{\dashv\mu\nu}\!=\!(1/N)\partial N/\partial N_{\nu\mu}$
appeared in the field equations but was not a genuine inverse of $N_{\nu\mu}$.
In addition, it would be impossible to derive the
$\Lambda_z=0$, ${\mathcal L}_m\!=\!0$ version of the theory from a purely affine
Lagrangian density, thus removing a motivation for the theory.
Note that we also cannot use the definition $N\!=\!\rm{det}(tr(N_{\nu\mu}))\,$ as in \cite{Borchsenius76}
because $\rm{det}(tr(N_{\nu\mu}))$ and $\rm{det}(tr(\hhR_{\nu\mu}))$
would not depend on the traceless part of the fields.

\section{\label{NonsymmetricCase}The case $\hz^{\nu\mu}\!=\!0$ with nonsymmetric fields}
Let us consider the theory for the special case $\hz^{\mu\nu}\!=\!0$, or more precisely for
\begin{eqnarray}
\label{cnumber}
\tGam^\alpha_{\nu\mu}=tr(\tGam^\alpha_{\nu\mu})\Isigma/d,~~~\sg^{\nu\mu}=tr(\sg^{\nu\mu})\Isigma/d.
\end{eqnarray}
In this case $\Am_\nu$ and $\RmN\!N^{\dashv[\nu\mu]}$ are the only independent
variables in (\ref{JSlag5}) which are not just multiples of the identity matrix $\Isigma$.
This assumption is both coordinate independent and gauge independent,
considering (\ref{tGamtransform},\ref{NItransform}).
We assume this special case because it gives us Einstein-Maxwell-Yang-Mills theory,
and because it greatly simplifies the theory.
With the assumption (\ref{cnumber}) we also have $\tR_{\nu\mu}=tr(\tR_{\nu\mu})\Isigma/d$, and the term
$([\Am_\alpha,\tGam{^\alpha_{\nu\mu}}]-[\Am_{(\nu},\tGam^\alpha_{\mu)\alpha}])\,i\sqrt{\sixteenpid\Lambda_b}\,$
vanishes in the Lagrangian density (\ref{JSlag5}).
It is important to emphasize that any solution of the restricted theory (\ref{cnumber}) will also
be a solution of the more general theory.

Setting $\delta{\mathcal L}/\delta\Am_\tau\!=\!0$ and using the definition (\ref{fdef2})
of $f^{\nu\mu}$ gives the ordinary Maxwell-Yang-Mills equivalent of Ampere's law,
\begin{eqnarray}
\label{Ampere2}
\fl (\Rmg  f^{\omega\tau})_{,\,\omega}-i\sqrt{\sixteenpid\Lambda_b}\,\Rmg  [f^{\omega\tau}\!,\Am_\omega]=\nofourpi \Rmg  j^\tau,
\end{eqnarray}
where the source current $j^\tau$ is defined by
\begin{eqnarray}
\label{jdef2}
j^\tau=\frac{-1}{\Rmg}\frac{\delta {\mathcal L}_m}{\delta \Am_\tau}\,.
\end{eqnarray}
Setting $\delta{\mathcal L}/\delta\tGam^\beta_{\tau\rho}\!=\!0$ using a Lagrange multiplier term
$tr[\Omega^\rho\tGam^\alpha_{\![\alpha\rho]}]$ to enforce the symmetry (\ref{JScontractionsymmetric2}),
and using the result $tr[(\Rmg  f^{\omega\tau})_{,\,\omega}]=\nofourpi\Rmg tr[j^\tau]$
derived from (\ref{Ampere2},\ref{Amsplit},\ref{tauproperties}) gives the connection equations,
\begin{eqnarray}
\label{JSconnection2}
\fl tr[(\RmN\!N^{\dashv\rho\tau})_{\!,\,\beta}
+\tGam^\tau_{\sigma\beta}\RmN\!N^{\dashv\rho\sigma}
+\tGam^\rho_{\beta\sigma}\RmN\!N^{\dashv\sigma\tau}
-\tGam^\alpha_{\beta\alpha}\RmN\!N^{\dashv\rho\tau}]\nonumber~~~~~~~\\
\fl ~~~~~~~~~~~~~~~~~~~~~~~~~~~~~~~~~~~~~~~~~~~~~~~~~~~~~
=\frac{ i4\pi\sqrt{2}}{(n\!-\!1)\Lambda_b^{1/2}}\,\Rmg\, tr[j^{[\rho}]\delta^{\tau]}_\beta.~~~~~~~
\end{eqnarray}
Setting $\delta{\mathcal L}/\delta(\RmN\!N^{\dashv(\mu\nu)})\!=\!0$
using the identities $\RmN\!\nobreak=\nobreak\![\pm det(\RmN\!N^{\dashv\mu\nu})]^{1/d(n-2)}$ and
$\Rmg\!=\![\pm det(\RmN\!N^{\dashv(\mu\nu)})]^{1/d(n-2)}$ gives our equivalent of the Einstein equations,
\begin{eqnarray}
\label{JSEinstein2}
\fl\frac{\lower1pt\hbox{$1$}}{d}\,tr[\tR_{(\nu\mu)}
+\Lambda_bN_{(\nu\mu)}+\Lambda_z\sg_{\nu\mu}]=8\pi tr[S_{\nu\mu}],
\end{eqnarray}
where $S_{\nu\mu}$ is defined by
\begin{eqnarray}
\label{Sdef2}
S_{\nu\mu}\!&\equiv&2\frac{\delta{\mathcal L}_m}{\delta(\RmN\!N^{(\mu\nu)})}
=2\frac{\delta{\mathcal L}_m}{\delta(\Rmg  \sg^{\mu\nu})}.
\end{eqnarray}
Setting $\delta{\mathcal L}/\delta(\RmN\!N^{\dashv[\mu\nu]})\!\nobreak=\nobreak\!0$
using the identities $\RmN\!\nobreak=\nobreak\![\pm det(\RmN\!N^{\dashv\mu\nu})]^{1/d(n-2)}$ and
$\Rmg\!=\![\pm det(\RmN\!N^{\dashv(\mu\nu)})]^{1/d(n-2)}$ gives,
\begin{eqnarray}
\label{JSantisymmetric2}
\fl \tR_{[\nu\mu]}
\!+2\Am_{[\nu,\mu]}\,i\sqrt{\sixteenpid\Lambda_b}
\!+\sixteenpid\Lambda_b[\Am_\nu,\Am_\mu]
+\Lambda_bN_{[\nu\mu]}=0.
\end{eqnarray}
Note that the antisymmetric field equations (\ref{JSantisymmetric2}) lack a source term
because ${\mathcal L}_m$ in (\ref{JSlag5}) contains only
$\Rmg  \sg^{\mu\nu}\!=\!\RmN\!N^{\dashv(\nu\mu)}$ from (\ref{sgdef2}),
and not $\RmN\!N^{\dashv[\nu\mu]}$. The trace operations in
(\ref{JSconnection2},\ref{JSEinstein2}) occur because we are assuming the special
case (\ref{cnumber}). The off-diagonal matrix components of
$\delta{\mathcal L}/\delta\tGam^\beta_{\tau\rho}$ and
$\delta{\mathcal L}/\delta(\RmN\!N^{\dashv(\mu\nu)})$ vanish because
with (\ref{cnumber}), the Lagrangian density contains no off-diagonal matrix
components of $\tGam^\beta_{\tau\rho}$ and $\RmN\!N^{\dashv(\mu\nu)}$.
The trace operation sums up the contributions from the diagonal matrix components of
$\tGam^\beta_{\tau\rho}$ and $\RmN\!N^{\dashv(\mu\nu)}$ because
(\ref{cnumber}) means that for a given set of tensor indices,
all of the diagonal matrix components are really the same variable.

To put (\ref{Ampere2}-\ref{JSantisymmetric2}) into a form which looks more like the ordinary
Einstein-Maxwell-Yang-Mills field equations we need to do some preliminary calculations.
The definitions (\ref{sgdef2},\ref{fdef2}) of $\sg_{\nu\mu}$ and $f_{\nu\mu}$
can be inverted to give $N_{\nu\mu}$ in terms of $\sg_{\nu\mu}$ and $f_{\nu\mu}$.
An expansion in powers of $\Lambda_b^{\!-1}$ is derived in \Ax\ref{ApproximateFandg},
\begin{eqnarray}
\label{approximateNbar2}
\fl N_{(\nu\mu)}&\!\!=\!& \sg_{\nu\mu}-2\!\left(f^\sigma\!{_{(\nu}}f_{\mu)\sigma}
-\frac{1}{2(n\!-\!2)}\,\sg_{\nu\mu} \frac{tr(f^\rho\!{_\sigma}\!f^\sigma\!{_\rho})}{d}\right)\!\Lambda_b^{\!-1}
+(f^3)\Lambda_b^{\!-3/2}\dots~~~~~~\\
\label{approximateNhat2}
\fl N_{[\nu\mu]}&\!\!=\!& f_{\nu\mu}i\sqrt{\sixteenpid}\,\Lambda_b^{\!-1/2}
+(f^2)\Lambda_b^{\!-1}\dots.
\end{eqnarray}
Here $(f^3)\Lambda_b^{\!-3/2}$ and $(f^2)\Lambda_b^{\!-1}$ are terms like
$f^\rho\!{_\sigma}f^\sigma\!{_{(\mu}}f_{\nu)\rho}\Lambda_b^{\!-3/2}$ and
$f^\sigma\!{_{[\nu}}f_{\mu]\sigma}\Lambda_b^{\!-1}\!$.

Because of the assumption (\ref{cnumber}) and the trace operation in (\ref{JSconnection2}),
the connection equations (\ref{JSconnection2}) are the same as with
the Abelian theory\cite{xShifflett} but with the substitution of $tr[f_{\nu\mu}]/d$
and $tr[j^\nu]/d$ instead of $f_{\nu\mu}$ and $j^\nu$.
Therefore the solution of the connection equations
from \cite{xShifflett}
can again be abbreviated as
\begin{eqnarray}
\label{tGam}
\fl \tGam^\alpha_{(\nu\mu)}&=&I\Gamma^\alpha_{\nu\mu}+(f'f)\Lambda_b^{\!-1}\dots
\,~~~~~~\tGam^\alpha_{[\nu\mu]}=(f')\Lambda_b^{\!-1}\dots,
\end{eqnarray}
where $\Gamma^\alpha_{\nu\mu}$ is the Christoffel connection,
\begin{eqnarray}
\label{Christoffel}
\Gamma^\alpha_{\nu\mu}&=&\frac{\lower2pt\hbox{$1$}}{2}\,g^{\alpha\sigma}(g_{\mu\sigma,\nu}
+g_{\sigma\nu,\mu}-g_{\nu\mu,\sigma}).
\end{eqnarray}
Substituting (\ref{tGam}) using (\ref{Ricciaddition2}) shows that as in \cite{xShifflett},
the non-Abelian Ricci tensor (\ref{HermitianizedRiccit2}) can again be abbreviated as
\begin{eqnarray}
\label{tRapproxG2}
\fl \tR_{(\nu\mu)}&=&IR_{\nu\mu}\!+(f'f')\Lambda_b^{\!-1}+(ff'')\Lambda_b^{\!-1}\dots,
\label{tRapprox02}
~~~~~~~~\tR_{[\nu\mu]}=(f'')\Lambda_b^{\!-1/2}\dots,~~~~
\end{eqnarray}
where $R_{\nu\mu}\!=\!R_{\nu\mu}(\Gamma)$ is the ordinary Ricci tensor.
Here $(f'f')\Lambda_b^{\!-1}$, $(ff'')\Lambda_b^{\!-1}$ and $(f'')\Lambda_b^{\!-1/2}$
refer to terms like $tr(f^\sigma\!{_{\nu;\alpha}}\!)tr(f^\alpha\!{_{\mu;\sigma}}\!)\Lambda_b^{\!-1}\!$,
$tr(f^\alpha\!{_\tau}\!)tr(f^\tau\!\!{_{(\nu;\,\mu)}}{_{;\alpha}}\!)\Lambda_b^{\!-1}$ and
$tr(f_{[\nu\mu,\alpha];}{^\alpha})\Lambda_b^{\!-1/2}$.

Combining (\ref{approximateNbar2},\ref{tRapproxG2},\ref{Lambdadef}) with the
symmetric field equations (\ref{JSEinstein2}) and their contraction gives
\begin{eqnarray}
\label{JSEinstein6}
\fl G_{\nu\mu}
&\!=&8\pi\, tr(T_{\nu\mu})
+2\!\left(\frac{tr(f^\sigma\!{_{(\nu}}f_{\mu)\sigma})}{d}
-\frac{1}{4}\,g_{\nu\mu} \frac{tr(f^{\rho\sigma}\!f_{\sigma\rho})}{d}\right)\nonumber\\
\fl &&+\Lambda\left(\frac{n}{2}-1\right)g_{\nu\mu}
+(f^3)\Lambda_b^{\!-1/2}+(f'f')\Lambda_b^{\!-1}+(ff'')\Lambda_b^{\!-1}\dots,
\end{eqnarray}
where the Einstein tensor and energy-momentum tensor are
\begin{eqnarray}
\label{Gdef2}
\fl G_{\nu\mu}&=&R_{\nu\mu}-\frac{\lower1pt\hbox{$1$}}{2}g_{\nu\mu}\,R^\alpha_\alpha,~~~~~
\label{Tdef2}
T_{\nu\mu}=S_{\nu\mu}\!-\frac{\lower1pt\hbox{$1$}}{2}\,g_{\nu\mu}S^\alpha_\alpha.
\end{eqnarray}
Here $(f^3)\Lambda_b^{\!-1/2}\!$, $(f'\!f')\Lambda_b^{\!-1}$ and $(f\!f'')\Lambda_b^{\!-1}$ ara terms like
$tr(f^\rho\!{_\sigma}f^\sigma\!{_{(\mu}}f_{\nu)\rho})\Lambda_b^{\!-1/2}\!$,
$tr(f^\sigma{_{\nu;\alpha}})tr(f^\alpha{_{\mu;\sigma}})\Lambda_b^{\!-1}$
and $tr(f^{\alpha\tau})tr(f_{\tau(\nu;\,\mu)}{_{;\alpha}})\Lambda_b^{\!-1}$.
This shows that the Einstein equations (\ref{JSEinstein6}) match those of
Einstein-Maxwell-Yang-Mills theory except for extra terms which will be
very small relative to the leading order terms because of the
large value $\Lambda_b\!\sim\! 10^{63}cm^{-2}$ from (\ref{Lambdab}).

Combining (\ref{approximateNhat2},\ref{tRapprox02}) with the antisymmetric field equations (\ref{JSantisymmetric2}) gives
\begin{eqnarray}
\label{fapprox2}
\fl f_{\nu\mu}=2\Am_{[\mu,\nu]}
\!+i\sqrt{\sixteenpid\Lambda_b}\,[\Am_\nu,\Am_\mu]+(f^2)\Lambda_b^{\!-1/2}+(f'')\Lambda_b^{\!-1}\dots.
\end{eqnarray}
Here $(f^2)\Lambda_b^{\!-1/2}$ and $(f'')\Lambda_b^{\!-1}$ are terms like
$f^\sigma\!{_{[\nu}}f_{\mu]\sigma}\Lambda_b^{\!-1}$
and $tr(f_{[\nu\mu,\alpha];}{^\alpha})\Lambda_b^{\!-1/2}\!$.
From (\ref{Lambdab}) we see that the $f_{\nu\mu}$ in Ampere's law (\ref{Ampere2})
matches the Maxwell-Yang-Mills tensor (\ref{FYangMills}) except for extra terms
which will be very small relative to the leading order terms because of the
large value $\Lambda_b\!\sim\! 10^{63}cm^{-2}$ from (\ref{Lambdab}).

Let us do a quantitative comparison of the $\hz^{\nu\mu}\!=\!0$ case to Einstein-Maxwell-Yang-Mills theory.
To do this we will consider the magnitude of the extra terms in
the Einstein equations and the Maxwell-Yang-Mills field tensor for worst-case field strengths and
rates-of-change accessible to measurement, and compare these to the ordinary terms.
In particular we will evaluate extra terms in the Einstein equations (\ref{JSEinstein6})
like $tr(f^\rho{_\sigma}f^\sigma\!{_{(\mu}}f_{\nu)\rho})\Lambda_b^{\!-1/2}$,
$tr(f^\sigma{_{\nu;\alpha}})tr(f^\alpha{_{\mu;\sigma}})\Lambda_b^{\!-1}$ and
$tr(f^{\alpha\tau})tr(f_{\tau(\nu;\,\mu)}{_{;\alpha}})\Lambda_b^{\!-1}$
and compare these to the ordinary Maxwell-Yang-Mills term.
Likewise we will evaluate extra terms in the Maxwell-Yang-Mills field tensor
(\ref{fapprox2}) like $f^\sigma\!{_{[\nu}}f_{\mu]\sigma}\Lambda_b^{\!-1/2}$
and $tr(f_{[\nu\mu,\alpha];}{^\alpha})\Lambda_b^{\!-1}$ and compare these
to $f_{\mu\nu}$ which appears in Ampere's law (\ref{Ampere2}).

We assume that the worst-case field strengths and rates of change
accessible to measurement will be purely electromagnetic fields.
Also, because we will just be doing order-of-magnitude calculations, we will neglect
mixing in $f_{\mu\nu}$ and we will use the electromagnetic coupling constant.
In geometrized units with the Heaviside-Lorentz convention
an elementary charge has $e\!=\!1.38\times\! 10^{-34}cm$.
If we assume that charged particles retain $f^1{_0}\!\sim\!e/\fourpi r^2$
down to the smallest radii probed by particle physics
experiments ($10^{-17}{\rm cm}$) we have from (\ref{Lambdab}),
\begin{eqnarray}
\label{highenergyskew}
|f^1{_0}|\Lambda_b^{\!-1/2}&\sim& \Lambda_b^{\!-1/2}e/\fourpi (10^{-17})^2\sim 10^{-31},\\
\label{highenergyderiv1}
|f^1{_{0;1}}/f^1{_0}|^2\Lambda_b^{\!-1}&\sim& 4\Lambda_b^{\!-1}/(10^{-17})^2\sim 10^{-29},\\
\label{highenergyderiv2}
|f^1{_{0;1;1}}/f^1{_0}|\Lambda_b^{\!-1}&\sim& 6\Lambda_b^{\!-1}/(10^{-17})^2\sim 10^{-29}.
\end{eqnarray}
The fields at $10^{-17}{\rm cm}$ from an elementary charge
would be larger than near any macroscopic charged object.
Here $f^1{_0}$ is assumed to be in some
standard spherical or cartesian coordinate system. If an equation has a tensor term which can
be neglected in one coordinate system, it can be neglected in any coordinate system,
so it is only necessary to prove it in one coordinate system.
So for electric monopole fields, the extra terms in the Einstein equations (\ref{JSEinstein6})
must be $<\!10^{-29}$ of the ordinary Maxwell-Yang-Mills term.
Similarly the extra terms
in the Maxwell-Yang-Mills field tensor (\ref{fapprox2}) must be $<\!10^{-29}$ of $f_{\nu\mu}$.
Also, for the highest energy electromagnetic waves known in nature ($10^{20}$eV, $10^{34}$Hz)
we have from (\ref{Lambdab}),
\begin{eqnarray}
\label{gammaderiv1}
|f^1{_{0;1}}/f^1{_0}|^2\Lambda_b^{\!-1}&\sim& (E/\hbar c)^2\Lambda_b^{\!-1}\sim 10^{-13},\\
\label{gammaderiv2}
|f^1{_{0;1;1}}/f^1{_0}|\Lambda_b^{\!-1}&\sim& (E/\hbar c)^2\Lambda_b^{\!-1}\sim 10^{-13}.
\end{eqnarray}
So for electromagnetic waves, the extra terms in the Einstein equations (\ref{JSEinstein6})
must be $<\!10^{-13}$ of the ordinary Maxwell-Yang-Mills term.
Similarly the extra terms in the Maxwell-Yang-Mills field tensor (\ref{fapprox2}) must be
$<\nobreak\!10^{-13}$ of $f_{\mu\nu}$ which appears in Ampere's law (\ref{Ampere2}).

From this analysis we see that these extra terms in the field equations
(\ref{JSEinstein6},\ref{fapprox2},\ref{Ampere2}) are far below
the level that could be detected by experiment for worst-case field strengths and rates of
change accessible to measurement. At least we have made great efforts
to find an experiment in which these extra terms would be evident,
and we have been unable to find such an experiment.
As shown in \cite{xShifflett}, the ordinary Lorentz force equation can be derived from the
divergence of the Einstein equations for the purely electromagnetic case of this theory.
In \cite{xShifflett} we also presented an exact electromagnetic plane-wave solution
which is identical to its counterpart in Einstein-Maxwell theory.
And in \cite{xShifflett} we presented an exact electric monopole solution which matches the
Reissner-Nordstr\"{o}m solution except for additional terms which are $<10^{-65}$ of
the usual terms for worst-case radii accessible to measurement.

We wish to emphasize that the ${\mathcal L}_m$ term in (\ref{JSlag4}) allows coupling to
additional fields via a symmetric metric $g_{\mu\nu}$ and Hermitian vector potential $\Am_\mu$,
just as in Einstein-Maxwell-Yang-Mills theory.
Our ${\mathcal L}_m$ can contain the same fermion and Higgs field terms as
in Weinberg-Salam theory or flipped $SU\!(5)$ GUT theory.
And when we do this we will get the same
energy-momentum tensor (\ref{Sdef2},\ref{Tdef2}) in the Einstein equations (\ref{JSEinstein6}),
and the same current (\ref{jdef2}) in the Maxwell-Yang-Mills equivalent of
Ampere's law (\ref{Ampere2},\ref{fapprox2}). In addition, the equations of motion of fermion and Higgs fields
will be unchanged, and the components of $\Am_\mu$ will mix and acquire mass in the usual way
(the $\Am_\mu$ mass terms will get lumped into $j^\tau$ in (\ref{Ampere2},\ref{jdef2})).

One aspect of this theory which might differ from
Einstein-Maxwell-Yang-Mills theory is discussed in detail at the end of section 5
of \cite{xShifflett} for the purely electromagnetic case,
although it is unclear whether it is really a difference or not.
To see what this is we take the curl of (\ref{fapprox2}),
in which case the $2\Am_{[\mu,\nu]}$ term falls out,
and from the $f_{\nu\mu}$ and $(f'')\Lambda_b^{\!-1}$ terms we get\cite{xShifflett},
\begin{eqnarray}
\label{Proca}
f_{[\nu\mu,\alpha]}=(-f_{[\nu\mu,\alpha];}{^\sigma}{_{;\sigma}}+{\rm apparently~negligible~terms})/2\Lambda_b\dots.~~~
\end{eqnarray}
This is similar to the Proca equation with the field
$\theta^\tau\!=\!\epsilon^{\tau\nu\mu\alpha}f_{[\nu\mu,\alpha]}/4$.
It suggests that the theory may allow $\theta^\tau$ Proca waves
with mass from (\ref{Proca},\ref{Lambdab}) close to the Planck mass.
For $d\!=\!2$ and $g_c\!=\!e/sin\theta_w$ we get
\begin{eqnarray}
\label{MProca}
\omega_{Proca}\!=\!\sqrt{2\Lambda_b}
\!=\!\frac{1}{l_P}\frac{g_c}{e}\sqrt{\frac{\alpha}{2d}},~~~~~
M_{Proca}\!=\!\hbar\omega_{Proca}\!=\!1.1\times\! 10^{18} GeV.~~~~
\end{eqnarray}
Using a Newman-Penrose $1/r$ expansion of the field equations we have shown that
continuous-wave solutions like $\theta^\tau\!\approx\!\epsilon^\tau sin(kr\!-\!\omega t)/r$ do not
exist in the theory\cite{ShifflettThesis}, but it is still possible that wave-packet solutions could exist.
If wave-packet $\theta^\tau$ solutions do occur, a calculation in \cite{xShifflett}
also suggests that they might have negative energy, although this calculation
is really based on the assumption that $\theta^\tau\!\approx\!\epsilon^\tau sin(kr\!-\!\omega t)/r$
solutions exist, and some questionable assumptions about terms being negligible.
If wave-packet $\theta^\tau$ solutions do exist,
and if they do have negative energy, there is still a possible interpretation of
the $\theta^\tau$ field as a built-in Pauli-Villars field,
with a cutoff mass (\ref{MProca}) which is
close to $M_{Planck}=1.22\times\! 10^{19}GeV$ commonly assumed for this purpose.

The additional cosmological constant $\Lambda_z$ in our Lagrangian density
(\ref{JSlag4},\ref{JSlag5}) could have several contributions.
If there was a contribution from zero-point fluctuations it would
be approximately\cite{Zeldovich,Sahni,Peskin,Carroll}
\begin{eqnarray}
\label{Lambdaz}
\Lambda_{z0}=-\frac{\lower2pt\hbox{$\omega_c^4 l_P^2$}}{2\pi}\!\left({{\lower2pt\hbox{fermion}}\atop{\raise2pt\hbox{spin~states}}}
\!-\!{{\lower2pt\hbox{boson}}\atop{\raise2pt\hbox{spin~states}}}\right),
\end{eqnarray}
where $\omega_c$ is a cutoff frequency and $l_P=({\rm Planck~length})$.
Assuming the Pauli-Villars ghost idea discussed above,
$\omega_c\!=\!\omega_{Proca}$ from (\ref{MProca}),
$\Lambda_z\!\approx-\!\Lambda_b$ from (\ref{Lambdab}),
$d\!=\!2$, $g_c\!=\!e/sin\theta_w$,
and the particles of the Standard Model gives
\begin{eqnarray}
\label{zeropointLambda}
\frac{\Lambda_{z0}}{\Lambda_z}
=\frac{\lower2pt\hbox{$\omega_{Proca}^2 l_P^2$}}{\pi}(96-28)
=\frac{\alpha}{2d\pi}\left(\frac{g_c}{e}\right)^2(96-28)=.17.
\end{eqnarray}
So by this calculation, zero-point fluctuations would only contribute about
$17\%$ of $\Lambda_z$. Additional contributions to $\Lambda_z$ could perhaps
come from Higgs field vacuum energy and additional unknown fields.
It is unclear how this calculation would work out for flipped $SU\!(5)$ GUT theory.
Note that $\omega_{Proca}$ and $\Lambda_b$ from (\ref{MProca}) depend on the
coupling constant $g_c$, so their values should ``run'' with frequency.
The contribution to $\Lambda_z$ from zero-point fluctuations (\ref{zeropointLambda})
would be slightly modified if we used a ``bare'' $g_c$
calculated at $\omega_{Proca}$ instead of a low energy value.
Also note that the Pauli-Villars ghost idea might not be necessary or correct,
in which case we could make (\ref{zeropointLambda})
closer to $100\%$ by assuming a slightly larger $\omega_c$.

\section{\label{SymmetricCase}The case $|\hz^{\nu\mu}|\!\ll\!1$ with symmetric fields}
This theory definitely differs from Einstein-Maxwell-Yang-Mills theory in that the symmetric fields can be non-Abelian,
with traceless components. To investigate this let us
calculate the field equations with the Lagrangian density (\ref{JSlag5},\ref{HermitianizedRiccit2})
and the special case $\Am_\nu\!=\!0$, $N^{\dashv[\mu\nu]}\!=\!0$, $\tGam^\alpha_{[\nu\mu]}\!=\!0$.
Setting $\delta{\mathcal L}/\delta(\RmN\!N^{\dashv(\mu\nu)})\!=\!0$ and
using $\RmN\!\nobreak=\nobreak\!\Rmg\!=\![\pm det(\RmN\!N^{\dashv(\mu\nu)})]^{1/d(n-2)}$
gives our equivalent of the Einstein equations,
\begin{eqnarray}
\label{JSEinstein3}
\fl \frac{\lower1pt\hbox{$1$}}{d}\,(\tR_{(\nu\mu)}+\Lambda\sg_{\nu\mu})=8\pi S_{\nu\mu}~~~~~~~\\
\label{Sdef3}
\fl {\rm where~~~} S_{\nu\mu}\!\equiv 2\frac{\delta{\mathcal L}_m}{\delta(\RmN\!N^{(\mu\nu)})}
=2\frac{\delta{\mathcal L}_m}{\delta(\Rmg  \sg^{\mu\nu})}\,.
\end{eqnarray}
For present purposes we assume $S_{\nu\mu}\!=\!0$ and $\Lambda\!=\!0$.
Setting $\delta{\mathcal L}/\delta\tGam^\beta_{\tau\rho}\!=\!0$
using $\tGam^\alpha_{[\nu\mu]}\!=\!0$, $N^{\dashv[\mu\nu]}\!=\!0$, $\Am_\nu\!=\!0$
and (\ref{sgdef2}) gives the connection equations\cite{ShifflettThesis},
\begin{eqnarray}
\label{JSconnection2rewrite}
\fl&&\!\!\!(\Rmg\sg^{\rho\tau})_{\!,\,\beta}
+\frac{\lower2pt\hbox{$1$}}{2}\tGam^\rho_{\beta\mu}\Rmg\sg^{\mu\tau}
+\frac{\lower2pt\hbox{$1$}}{2}\Rmg\sg^{\rho\nu}\tGam^\tau_{\nu\beta}
+\frac{\lower2pt\hbox{$1$}}{2}\tGam^\tau_{\beta\mu}\Rmg\sg^{\mu\rho}
+\frac{\lower2pt\hbox{$1$}}{2}\Rmg  \sg^{\tau\nu}\tGam^\rho_{\nu\beta}\nonumber\\
\fl &&~~~~~~~~~~~~~~~~~~~~~~~~~~~~~~~~~~~~~~~~~~
-\frac{\lower2pt\hbox{$1$}}{2}\tGam^\alpha_{\beta\alpha}\Rmg\sg^{\rho\tau}
-\frac{\lower2pt\hbox{$1$}}{2}\Rmg\sg^{\rho\tau}\tGam^\alpha_{\beta\alpha}=0.
\end{eqnarray}

We will only consider the case where the traceless components are small,
similar to linearized gravity,
\begin{eqnarray}
|\hz^{\nu\mu}|\!\ll\!1,~~~
|H^\alpha_{\nu\mu}|\!\ll\!|\Gamma^\alpha_{\nu\mu}|,~~~
\tGam^\alpha_{\nu\mu}\!=\!I\Gamma^\alpha_{\nu\mu}\!+\! H^\alpha_{\nu\mu}+{\mathcal O}(\hz^2).
\end{eqnarray}
Here $\hz^{\nu\mu}$ is defined in (\ref{sgdef2}) and
$\Gamma^\alpha_{\nu\mu}$ is the Christoffel connection (\ref{Christoffel})
formed from the physical metric $g_{\nu\mu}$ with no traceless components.
The connection equations (\ref{JSconnection2rewrite}) to ${\mathcal O}(\hz)$ are
\begin{eqnarray}
\fl&&\!\!\!(\rmg\,\hz^{\rho\tau})_{\!,\,\beta}
+\!\frac{\lower2pt\hbox{$1$}}{2}\Gamma^\rho_{\beta\mu}\rmg\,\hz^{\mu\tau}
\!\!-\frac{\lower2pt\hbox{$1$}}{2}H^\rho_{\beta\mu}\rmg\,g^{\mu\tau}
\!+\!\frac{\lower2pt\hbox{$1$}}{2}\rmg\,\hz^{\rho\nu}\!\Gamma^\tau_{\nu\beta}
\!-\frac{\lower2pt\hbox{$1$}}{2}\rmg\,g^{\rho\nu}H^\tau_{\nu\beta}\nonumber\\
\fl &&\!\!\!+\frac{\lower2pt\hbox{$1$}}{2}\Gamma^\tau_{\beta\mu}\rmg\,\hz^{\mu\rho}
-\frac{\lower2pt\hbox{$1$}}{2}H^\tau_{\beta\mu}\rmg\,g^{\mu\rho}
+\frac{\lower2pt\hbox{$1$}}{2}\rmg\,\hz^{\tau\nu}\Gamma^\rho_{\nu\beta}
-\frac{\lower2pt\hbox{$1$}}{2}\rmg\,g^{\tau\nu}H^\rho_{\nu\beta}\nonumber\\
\fl &&\!\!\!-\frac{\lower2pt\hbox{$1$}}{2}\Gamma^\alpha_{\beta\alpha}\rmg\,\hz^{\rho\tau}
+\frac{\lower2pt\hbox{$1$}}{2}H^\alpha_{\beta\alpha}\rmg\,g^{\rho\tau}
\!-\frac{\lower2pt\hbox{$1$}}{2}\rmg\,\hz^{\rho\tau}\Gamma^\alpha_{\beta\alpha}
+\frac{\lower2pt\hbox{$1$}}{2}\rmg\,g^{\rho\tau}H^\alpha_{\beta\alpha}=0.~~~
\end{eqnarray}
Using $(\rmg\,)_{\!,\,\beta}=\rmg\,\Gamma^\alpha_{\beta\alpha}$ and dividing by $\rmg\,$ gives
\begin{eqnarray}
\fl 0&=&\hz^{\rho\tau}{_{\!;\,\beta}}
-H^\rho_{\beta\mu}g^{\mu\tau}
-g^{\rho\nu}H^\tau_{\nu\beta}+g^{\rho\tau}H^\alpha_{\beta\alpha}.
\end{eqnarray}
Combining the permutations of this gives
\begin{eqnarray}
\fl 0&=&(\hz_{\omega\lambda;\beta}
-H_{\omega\beta\lambda}
-H_{\lambda\omega\beta}
+g_{\omega\lambda}H^\alpha_{\beta\alpha})\nonumber\\
\fl &-&(\hz_{\beta\omega;\lambda}
-H_{\beta\lambda\omega}
-H_{\omega\beta\lambda}
+g_{\beta\omega}H^\alpha_{\lambda\alpha})\nonumber\\
\fl &-&(\hz_{\lambda\beta;\omega}
-H_{\lambda\omega\beta}
-H_{\beta\lambda\omega}
+g_{\lambda\beta}H^\alpha_{\omega\alpha})\\
\fl &=&2H_{\beta\lambda\omega}
+\hz_{\omega\lambda;\beta}
-\hz_{\beta\omega;\lambda}
-\hz_{\lambda\beta;\omega}
+g_{\omega\lambda}H^\alpha_{\beta\alpha}
-g_{\beta\omega}H^\alpha_{\lambda\alpha}
-g_{\lambda\beta}H^\alpha_{\omega\alpha}.~~~
\end{eqnarray}
Contracting this with $g^{\beta\omega}$
gives
\begin{eqnarray}
\label{Hcontracted13}
\fl 0&=&2H^\omega_{\lambda\omega}-\hz^\omega_{\omega;\lambda}
-nH^\alpha_{\lambda\alpha}~~~~~~~~~~~
\Rightarrow~~~ H^\omega_{\lambda\omega}=\frac{\lower2pt\hbox{$1$}}{(2\!-\!n)}\hz^\omega_{\omega;\lambda}.~~~~~
\end{eqnarray}
So the ${\mathcal O}(\hz)$ solution of the connection equations (\ref{JSconnection2rewrite}) is
\begin{eqnarray}
\label{Hfinal2}
\fl H_{\alpha\nu\mu}
=\frac{\lower2pt\hbox{$1$}}{2}
(\hz_{\alpha\nu;\mu}
\!+\!\hz_{\mu\alpha;\nu}
\!-\!\hz_{\nu\mu;\alpha})
+\frac{\lower2pt\hbox{$1$}}{2(2\!-\!n)}(g_{\alpha\nu}\hz^\omega_{\omega;\mu}
\!+\!g_{\mu\alpha}\hz^\omega_{\omega;\nu}
\!-\!g_{\nu\mu}\hz^\omega_{\omega;\alpha}).~~~~
\end{eqnarray}
Assuming $\tGam^\alpha_{\nu\mu}\!=\!I\Gamma^\alpha_{\nu\mu}\!+\!H^\alpha_{\nu\mu}\!+\!K^\alpha_{\nu\mu}+{\mathcal O}(\hz^3)$
and using a similar method\cite{ShifflettThesis} gives the ${\mathcal O}(\hz^2)$ solution of
the connection equations (\ref{JSconnection2rewrite}),
\begin{eqnarray}
\label{Kfinal2}
\fl K_{\beta\tau\rho}
&=&\frac{1}{4(2\!-\!n)}\,[-g_{\rho\tau}(\hz^\omega_\nu\hz^\nu_\omega)_{,\beta}
+g_{\beta\rho}(\hz^\omega_\nu\hz^\nu_\omega)_{,\tau}
+g_{\tau\beta}(\hz^\omega_\nu\hz^\nu_\omega)_{,\rho}\nonumber\\
\fl &&~~~~~~~~~~~+g_{\rho\tau}(\hz^\omega_{\omega;\sigma}\hz^\sigma_\beta+\hz^\sigma_\beta \hz^\omega_{\omega;\sigma})
-\hz^\omega_{\omega;\beta}\hz_{\rho\tau}-\hz_{\rho\tau}\hz^\omega_{\omega;\beta}]\nonumber\\
\fl &+&\frac{\lower2pt\hbox{$1$}}{4}[(\hz_{\sigma\beta;\rho}-\hz_{\rho\sigma;\beta})\hz^\sigma_\tau
+\hz_\tau^\sigma (\hz_{\sigma\beta;\rho}-\hz_{\rho\sigma;\beta})\nonumber\\
\fl &&~~+(\hz_{\beta\sigma;\tau}\!-\hz_{\sigma\tau;\beta})\hz^\sigma_\rho
+\hz_\rho^\sigma (\hz_{\beta\sigma;\tau}\!-\hz_{\sigma\tau;\beta})
+\hz_{\rho\tau;\sigma}\hz^\sigma_\beta
+\hz_\beta^\sigma \hz_{\rho\tau;\sigma}].~~~~~~
\end{eqnarray}

The field equations for $\hz_{\nu\mu}$ are found by substituting
the ${\mathcal O}(\hz)$ solution (\ref{Hfinal2})
into the traceless part of the exact field equations (\ref{JSEinstein3}) and
using (\ref{Ricciaddition2},\ref{Hcontracted13},\ref{sgdef2})
\begin{eqnarray}
\fl 0&=&2\,[H^\alpha_{\nu\mu;\alpha}-H^\alpha_{\alpha(\nu;\mu)}]\\
\label{Einsteinlinear2}
\fl &=&-\hz_{\nu\mu;\alpha;}{^\alpha}
+2\hz_{\alpha(\nu;\mu);}{^\alpha}
+\frac{\lower2pt\hbox{$1$}}{(n\!-\!2)}
g_{\nu\mu}\hz^\tau_{\tau;\alpha;}{^\alpha}.
\end{eqnarray}
Contracting this equation gives
\begin{eqnarray}
\label{contractedEinsteinlinear}
\hz^\tau_{\tau;\alpha;}{^\alpha}=(2\!-\!n)\hz^\tau_{\alpha;\tau;}{^\alpha}.
\end{eqnarray}
So we can also write the field equations as
\begin{eqnarray}
\label{Einsteinlinear3}
\fl 0&=&-\hz_{\nu\mu;\alpha;}{^\alpha}
+2\hz_{\alpha(\nu;\mu);}{^\alpha}
-g_{\nu\mu}\hz^\tau_{\alpha;\tau}{^\alpha}.
\end{eqnarray}

Now let us assume that we can ignore the difference between covariant derivative and ordinary derivative.
In that case (\ref{Einsteinlinear2},\ref{Einsteinlinear3}) match the ``gauge independent''
field equations\cite{Misner} of linearized gravity, but with a non-Abelian $\hz_{\nu\mu}$.
In linearized gravity one often assumes the Lorentz gauge
\begin{eqnarray}
\label{Lorentzgauge}
\hz_{\nu\alpha;}{^\alpha}\!=\!0.
\end{eqnarray}
Here we do not have the same freedom because in the coordinate transformation
$x^\nu\!\rightarrow\!x^\nu\!+\!\xi^\nu$, $\hz_{\nu\mu}\!\rightarrow\!\hz_{\nu\mu}\!-\xi_{\nu;\mu}\!-\xi_{\mu;\nu}$
the parameter $\xi^\nu$ cannot be traceless like $\hz_{\nu\mu}$.
However, we can still seek solutions which satisfy (\ref{Lorentzgauge}).
Analogous with linearized gravity we have a z-directed plane-wave solution
\begin{eqnarray}
\label{linearsolution}
\fl \hz_{\nu\mu}\approx sin(\omega t-kz)\!
\begin{pmatrix}{0&0&0&0\cr 0&\hz_+&\hz_\times&0\cr 0&\hz_\times&-\hz_+&0\cr 0&0&0&0}\end{pmatrix},
~~~~~~k=\omega,
\end{eqnarray}
and a static spherically symmetric solution
\begin{eqnarray}
\label{blackholesolution}
\fl \hz_{\nu\mu}\approx
\begin{pmatrix}{4M/r\!\!&0&0&0\cr0&0&0&0\cr 0&0&0&0\cr 0&0&0&0}\end{pmatrix},
~~~~~~r=|\mathbf{x}-\mathbf{x}_p|.
\end{eqnarray}
Here $\hz_+,\hz_\times,M$ are constant traceless Hermitian matrices.
Note that (\ref{blackholesolution}) violates $|\hz^{\nu\mu}|\!\ll\!1$
near $r\!=\!0$, so a corresponding exact solution may not exist.

To find an effective energy-momentum tensor for $\hz_{\nu\mu}$
we extract the ${\mathcal O}(\hz^2)$ components\cite{ShifflettThesis} from the exact field equations (\ref{JSEinstein3})
using (\ref{Ricciaddition2},\ref{Hfinal2},\ref{Hcontracted13},\ref{Kfinal2},\ref{sgdef2}),
\begin{eqnarray}
\fl &&8\pi\tilde S_{\tau\rho}=-tr[K^\alpha_{\tau\rho;\alpha}
-K^\alpha_{\alpha(\tau;\rho)}
+H^\sigma_{\tau\rho}H^\alpha_{\sigma\alpha}
-H^\sigma_{\tau\alpha}H^\alpha_{\sigma\rho}]\\
\label{Sintermediate}
\fl&&=tr[-g_{\rho\tau}(\hz^\omega_\nu\hz^\nu_\omega)_{;\alpha;}{^\alpha}/4(n\!-\!2)
+g_{\rho\tau}(\hz^\omega_{\omega;\sigma}\hz^\sigma_\alpha)_{;}{^\alpha}/2(n\!-\!2)\nonumber\\
\fl &&~~~-(\hz^\omega_{\omega;\alpha}\hz_{\rho\tau})_{;}{^\alpha}/2(n\!-\!2)
-(\hz_{\sigma\alpha;(\rho}\hz^\sigma_{\tau)})_{;}{^\alpha}
+(\hz_{\rho\sigma}\hz^\sigma_\tau)_{;\alpha;}{^\alpha}/2
-(\hz_{\rho\tau;\sigma}\hz^\sigma_\alpha)_{;}{^\alpha}/2\nonumber\\
\fl &&~~~+\hz_{\tau\rho;}{^\sigma}\hz^\alpha_{\alpha;\sigma}/2(n\!-\!2)
-\hz^\omega_{\omega;\rho}\hz^\alpha_{\alpha;\tau}/4(n\!-\!2)
+\hz^\sigma_{\tau;\alpha}\hz^\alpha_{\rho;\sigma}/2
-\hz^\sigma_{\tau;\alpha}\hz_{\sigma\rho;}{^\alpha}/2\nonumber\\
\fl &&~~~+\hz^\sigma_{\alpha;\tau}\hz^\alpha_{\sigma;\rho}/4].
\end{eqnarray}
So the effective energy-momentum tensor is\cite{ShifflettThesis}
\begin{eqnarray}
\fl&&8\pi\tilde T_{\tau\rho}=8\pi \left(\tilde S_{\tau\rho}-\frac{\lower1pt\hbox{$1$}}{2}g_{\tau\rho}\tilde S^\mu_\mu\right)\\
\label{Tintermediate}
\fl&&=tr[-g_{\rho\tau}(\hz^\omega_\nu\hz^\nu_\omega)_{;\alpha;}{^\alpha}/8
+g_{\rho\tau}(\hz^\omega_\omega\hz^\mu_\mu)_{;\alpha;}{^\alpha}/8(n\!-\!2)
+g_{\rho\tau}(\hz^\sigma_{\alpha;\mu}\hz^\mu_\sigma)_{;}{^\alpha}/2\nonumber\\
\fl &&~~~-g_{\rho\tau}\hz^\mu_{\mu;}{^\sigma}\hz^\alpha_{\alpha;\sigma}/8(n\!-\!2)
-g_{\rho\tau}\hz^{\sigma\mu}{_{;\alpha}}\hz^\alpha_{\mu;\sigma}/4
+g_{\rho\tau}\hz^\sigma_{\mu;\alpha}\hz^\mu_{\sigma;}{^\alpha}/8\nonumber\\
\fl &&~~~-(\hz^\omega_{\omega;\alpha}\hz_{\rho\tau})_{;}{^\alpha}/2(n\!-\!2)
-(\hz_{\sigma\alpha;(\rho}\hz^\sigma_{\tau)})_{;}{^\alpha}
+(\hz_{\rho\sigma}\hz^\sigma_\tau)_{;\alpha;}{^\alpha}/2
-(\hz_{\rho\tau;\sigma}\hz^\sigma_\alpha)_{;}{^\alpha}/2\nonumber\\
\fl &&~~~+\hz_{\tau\rho;}{^\sigma}\hz^\alpha_{\alpha;\sigma}/2(n\!-\!2)
-\hz^\omega_{\omega;\rho}\hz^\alpha_{\alpha;\tau}/4(n\!-\!2)
+\hz^\sigma_{\tau;\alpha}\hz^\alpha_{\rho;\sigma}/2
-\hz^\sigma_{\tau;\alpha}\hz_{\sigma\rho;}{^\alpha}/2\nonumber\\
\fl &&~~~+\hz^\sigma_{\alpha;\tau}\hz^\alpha_{\sigma;\rho}/4].
\end{eqnarray}
From the field equations (\ref{Einsteinlinear2}) we get
\begin{eqnarray}
\fl 0&=&tr[(-\hz_{\nu\rho;}{^\alpha}{_{;\alpha}}+2\hz^\alpha_{(\nu;\rho);\alpha}
+g_{\rho\nu}\hz^\omega_{\omega;\alpha;}{^\alpha}/(n\!-\!2))\hz^\nu_\tau]\\
\fl &=&tr[-\hz_{\nu\rho;}{^\alpha}\hz^\nu_\tau
+\hz^\alpha_{\nu;\rho}\hz^\nu_\tau
+\hz^\alpha_{\rho;\nu}\hz^\nu_\tau
+\hz^\omega_{\omega;}{^\alpha}\hz_{\rho\tau}/(n\!-\!2)]_{;\alpha}\nonumber\\
\label{FromFieldEquations}
\fl &-&tr[-\hz_{\nu\rho;}{^\alpha}\hz^\nu_{\tau;\alpha}
+\hz^\alpha_{\nu;\rho}\hz^\nu_{\tau;\alpha}
+\hz^\alpha_{\rho;\nu}\hz^\nu_{\tau;\alpha}
+\hz^\omega_{\omega;}{^\alpha}\hz_{\rho\tau;\alpha}/(n\!-\!2)].
\end{eqnarray}
Using $tr(M_1M_2)\!=\!tr(M_2M_1)$,
the symmetrization and contraction of (\ref{FromFieldEquations}) are
\begin{eqnarray}
\label{Symmetrizedterm}
\fl 0&=&tr[-(\hz_{\nu\rho}\hz^\nu_\tau)_;{^\alpha}/2
+\hz^\alpha_{\nu;(\rho}\hz^\nu_{\tau)}
+\hz^\nu_{(\tau}\hz^\alpha_{\rho);\nu}
+\hz^\omega_{\omega;}{^\alpha}\hz_{\rho\tau}/(n\!-\!2)]_{;\alpha}\nonumber\\
\fl &-&tr[-\hz_{\nu\rho;}{^\alpha}\hz^\nu_{\tau;\alpha}
+\hz^\alpha_{\nu;(\rho}\hz^\nu_{\tau);\alpha}
+\hz^\alpha_{\rho;\nu}\hz^\nu_{\tau;\alpha}
+\hz^\omega_{\omega;}{^\alpha}\hz_{\rho\tau;\alpha}/(n\!-\!2)],~~~~~~~\\
\label{Symmetrizedtermcontracted}
\fl 0&=&tr[-(\hz^\sigma_\nu\hz^\nu_\sigma)_;{^\alpha}/2
+2\hz^\alpha_{\nu;\sigma}\hz^{\nu\sigma}
+(\hz^\omega_{\omega}\hz^\sigma_\sigma)_;{^\alpha}/2(n\!-\!2)]_{;\alpha}\nonumber\\
\fl &-&tr[-\hz^\nu_{\sigma;}{^\alpha}\hz^\sigma_{\nu;\alpha}
+2\hz^\alpha_{\nu;\sigma}\hz^{\nu\sigma}{_{;\alpha}}
+\hz^\omega_{\omega;}{^\alpha}\hz^\sigma_{\sigma;\alpha}/(n\!-\!2)].
\end{eqnarray}
Adding to (\ref{Tintermediate}) the expression
(\ref{Symmetrizedterm})/2$-g_{\rho\tau}$(\ref{Symmetrizedtermcontracted})$/8$
gives a simpler form of the effective energy-momentum tensor\cite{ShifflettThesis}
which is valid when $S_{\nu\mu}\!=0$ in (\ref{JSEinstein3}),
\begin{eqnarray}
\label{Ttilde}
\fl 8\pi\tilde T_{\tau\rho}&=&tr[-g_{\rho\tau}(\hz^\omega_\nu\hz^\nu_\omega)_{;\alpha;}{^\alpha}\!/16
+g_{\rho\tau}(\hz^\alpha_{\nu;\sigma}\hz^{\nu\sigma})_{;\alpha}/4
+g_{\rho\tau}(\hz^\omega_\omega\hz^\sigma_\sigma)_{;\alpha;}{^\alpha}\!/16(n\!-\!2)\nonumber\\
\fl &&~~~-(\hz_{\sigma\alpha;(\rho}\hz^\sigma_{\tau)})_{;}{^\alpha}
+(\hz_{\rho\sigma}\hz^\sigma_\tau)_{;\alpha;}{^\alpha}\!/4
-(\hz_{\rho\tau;\sigma}\hz^\sigma_\alpha)_{;}{^\alpha}\!/2
+(\hz^\nu_{(\tau}\hz^\alpha_{\rho);\nu})_{;}{^\alpha}\!/2\nonumber\\
\fl &&~~~+\hz^\sigma_{\alpha;\tau}\hz^\alpha_{\sigma;\rho}/4
-\hz^\omega_{\omega;\rho}\hz^\alpha_{\alpha;\tau}/4(n\!-\!2)
+\hz^\alpha_{\nu;(\rho|;\alpha}\hz^\nu_{|\tau)}/2].
\end{eqnarray}
Averaging over space or time, covariant derivatives commute
and gradients do not contribute\cite{Misner},
so the averaged effective energy-momentum tensor is
\begin{eqnarray}
\label{Ttildeaverage}
\fl 8\pi\!<\!\tilde T_{\tau\rho}\!\!>\!
&=&<tr[\,\hz^\sigma_{\alpha;\tau}\hz^\alpha_{\sigma;\rho}/4
-\hz^\omega_{\omega;\rho}\hz^\alpha_{\alpha;\tau}/4(n\!-\!2)
+\hz^\alpha_{\nu;\alpha;(\rho}\hz^\nu_{\tau)}/2]>.
\end{eqnarray}
This result is the same as
for gravitational waves\cite{Misner} but with a non-Abelian $\hz_{\nu\mu}$.
From (\ref{Ttildeaverage},\ref{Lorentzgauge},\ref{tauproperties}) we see that
the solution (\ref{linearsolution}) has positive energy density,
\begin{eqnarray}
\fl 8\pi\!<\!\tilde T_{00}\!\!>\!
&=&<\!tr[\hz^\sigma_{\alpha;0}\hz^\alpha_{\sigma;0}]\!>\!/4\\
&=&<tr[\hz^1_{1;0}\hz^1_{1;0}\!+\hz^1_{2;0}\hz^2_{1;0}\!+\hz^2_{1;0}\hz^1_{2;0}\!+\hz^2_{2;0}\hz^2_{2;0}]/4>\\
&=&tr[\hz_+^2+\hz_\times^2]\omega^2/4>0.
\end{eqnarray}

While solutions like (\ref{linearsolution},\ref{blackholesolution})
have not been observed, one must remember that
gravitational waves and black holes have not been observed directly either.
Solutions like (\ref{linearsolution},\ref{blackholesolution}) do not rule out the theory.
In fact if there is an exact solution corresponding to (\ref{blackholesolution}),
it might be a possible dark matter candidate.

\section{\label{Conclusions}Conclusions}

The Einstein-Schr\"{o}dinger theory is modified to include a
cosmological constant $\Lambda_z$ which multiplies the symmetric metric,
and by allowing the fields to be composed of Hermitian matrices.
The additional cosmological constant is assumed to be nearly cancelled by
Schr\"{o}dinger's ``bare'' cosmological constant $\Lambda_b$
which multiplies the nonsymmetric fundamental tensor, such that the total ``physical''
cosmological constant $\Lambda\!=\!\Lambda_b\!+\!\Lambda_z$ matches measurement.
If the symmetric part of the fields is assumed to be a multiple of the identity matrix,
the theory closely approximates Einstein-Maxwell-Yang-Mills theory.
The extra terms in the field equations all contain the
large constant $\Lambda_b\!\sim\!10^{63}cm^{-2}$ in the denominator,
and as a result these terms are $<10^{-13}$ of the usual terms
for worst-case fields and rates of change accessible to measurement.
Like Einstein-Maxwell-Yang-Mills theory, our theory is invariant under $U(1)$ and $SU(d)$
gauge transformations, and can be coupled to additional fields
using a symmetric metric and Hermitian vector potential.

\section{Acknowledgements}
This work was funded in part by the National Science Foundation under grant PHY~06-52448.

\appendix

\section{\label{hRproperties}Some properties of the non-Abelian Ricci tensor}
Substituting $\tGam^\alpha_{\nu\mu}\!=\!\Gamma^\alpha_{\nu\mu}
\!+\!\Upsilon^\alpha_{\nu\mu}$
into (\ref{HermitianizedRiccit2})
gives
\begin{eqnarray}
\fl \hR_{\nu\mu}(\tGam)
&=&\tGam^\alpha_{\nu\mu,\alpha}
-\tGam^\alpha_{\alpha(\nu,\mu)}
+\frac{\lower2pt\hbox{$1$}}{2}\tGam^\sigma_{\nu\mu}\tGam^\alpha_{\sigma\alpha}
+\frac{\lower2pt\hbox{$1$}}{2}\tGam^\alpha_{\sigma\alpha}\tGam^\sigma_{\nu\mu}
-\tGam^\sigma_{\nu\alpha}\tGam^\alpha_{\sigma\mu}\\
&=&(\Gamma^\alpha_{\nu\mu,\alpha}+\Upsilon^\alpha_{\nu\mu,\alpha})
-(\Gamma^\alpha_{\alpha(\nu,\mu)}+\Upsilon^\alpha_{\alpha(\nu,\mu)})
-(\Gamma^\sigma_{\nu\alpha}+\Upsilon^\sigma_{\nu\alpha})(\Gamma^\alpha_{\sigma\mu}+\Upsilon^\alpha_{\sigma\mu})\nonumber\\
&&+\frac{\lower2pt\hbox{$1$}}{2}(\Gamma^\sigma_{\nu\mu}+\Upsilon^\sigma_{\nu\mu})(\Gamma^\alpha_{\sigma\alpha}+\Upsilon^\alpha_{\sigma\alpha})
+\frac{\lower2pt\hbox{$1$}}{2}(\Gamma^\alpha_{\sigma\alpha}+\Upsilon^\alpha_{\sigma\alpha})(\Gamma^\sigma_{\nu\mu}+\Upsilon^\sigma_{\nu\mu})\\
&=&R_{\nu\mu}(\Gamma)+\Upsilon^\alpha_{\nu\mu,\alpha}-\Upsilon^\alpha_{\alpha(\nu,\mu)}
-\Gamma^\sigma_{\nu\alpha}\Upsilon^\alpha_{\sigma\mu}
-\Upsilon^\sigma_{\nu\alpha}\Gamma^\alpha_{\sigma\mu}
-\Upsilon^\sigma_{\nu\alpha}\Upsilon^\alpha_{\sigma\mu}\nonumber\\
&&+\frac{\lower2pt\hbox{$1$}}{2}(\Gamma^\sigma_{\nu\mu}\Upsilon^\alpha_{\sigma\alpha}
+\Upsilon^\sigma_{\nu\mu}\Gamma^\alpha_{\sigma\alpha}
+\Upsilon^\sigma_{\nu\mu}\Upsilon^\alpha_{\sigma\alpha}
+\Gamma^\alpha_{\sigma\alpha}\Upsilon^\sigma_{\nu\mu}
+\Upsilon^\alpha_{\sigma\alpha}\Gamma^\sigma_{\nu\mu}
+\Upsilon^\alpha_{\sigma\alpha}\Upsilon^\sigma_{\nu\mu})~~~~~~\\
\label{Ricciaddition2}
\fl \!&=&R_{\nu\mu}(\Gamma)+\Upsilon^\alpha_{\nu\mu;\alpha}
-\Upsilon^\alpha_{\alpha(\nu;\mu)}
-\Upsilon^\sigma_{\nu\alpha}\Upsilon^\alpha_{\sigma\mu}
+\frac{\lower2pt\hbox{$1$}}{2}\Upsilon^\sigma_{\nu\mu}\Upsilon^\alpha_{\sigma\alpha}
+\frac{\lower2pt\hbox{$1$}}{2}\Upsilon^\alpha_{\sigma\alpha}\Upsilon^\sigma_{\nu\mu}.
\end{eqnarray}

{
Substituting the $SU\!(d)$ gauge transformation
$\nGam{^\alpha_{\nu\mu}}\rightarrow`\nGam{^\alpha_{\nu\mu}}\!=\!U{\nGam}{^\alpha_{\nu\mu}}U^{-\!1}
+2\delta^\alpha_{[\nu} U_{,\mu]}U^{-\!1}$ from (\ref{gammatransform})
into $\hR_{\nu\mu}$ proves the result (\ref{hRtransform}),
and the result (\ref{gaugesymmetric2}) for a $U\!(1)$ gauge transformation
$\nGam^\alpha_{\rho\tau}\rightarrow\nGam^\alpha_{\rho\tau}\!-2iI\delta^\alpha_{[\rho}\varphi_{,\tau]}$
follows for the special case $U=Ie^{-i\varphi}$.
\begin{eqnarray}
\fl \hR_{\nu\mu}(`\nGam)
&=&\!`\nGam^\alpha_{\nu\mu,\alpha}
\!-`\nGam^\alpha_{\!(\alpha(\nu),\mu)}
\!+\frac{\lower2pt\hbox{$1$}}{2}`\nGam^\sigma_{\nu\mu}`\nGam^\alpha_{\!(\sigma\alpha)}
\!+\frac{\lower2pt\hbox{$1$}}{2}`\nGam^\alpha_{\!(\sigma\alpha)}`\nGam^\sigma_{\nu\mu}
\!-`\nGam^\sigma_{\nu\alpha}`\nGam^\alpha_{\sigma\mu}
\!-\frac{{`\nGam}{^\tau_{\![\tau\nu]}}{`\nGam}{^\rho_{\![\rho\mu]}}}{(n\!-\!1)}\\
&=&\!\!\left(U{\nGam}{^\alpha_{\nu\mu}}U^{-\!1}
+\delta^\alpha_\nu U_{,\mu}U^{-\!1}
-\delta^\alpha_\mu U_{,\nu}U^{-\!1}\right)\!\!{_{,\alpha}}\nonumber\\
\fl &-&\frac{\lower2pt\hbox{$1$}}{2}(U\nGam^\alpha_{\!(\alpha\nu)}U^{-\!1}){_{,\mu}}
-\frac{\lower2pt\hbox{$1$}}{2}(U\nGam^\alpha_{\!(\alpha\mu)}U^{-\!1}){_{,\nu}}\nonumber\\
\fl &+&\!\frac{\lower2pt\hbox{$1$}}{2}\!\left(U{\nGam}{^\sigma_{\nu\mu}}U^{-\!1}
+\delta^\sigma_\nu U_{,\mu}U^{-\!1}
-\delta^\sigma_\mu U_{,\nu}U^{-\!1}\right)\!U\nGam^\alpha_{\!(\sigma\alpha)}U^{-\!1}\nonumber\\
\fl &+&\!\frac{\lower2pt\hbox{$1$}}{2}U\nGam^\alpha_{\!(\sigma\alpha)}U^{-\!1}\!\left(U{\nGam}{^\sigma_{\nu\mu}}U^{-\!1}
+\delta^\sigma_\nu U_{,\mu}U^{-\!1}
-\delta^\sigma_\mu U_{,\nu}U^{-\!1}\right)\nonumber\\
\fl &-&\!\!\left(U{\nGam}{^\sigma_{\nu\alpha}}U^{-\!1}\!
+\delta^\sigma_\nu U_{,\alpha}U^{-\!1}\!
-\delta^\sigma_\alpha U_{,\nu}U^{-\!1}\! \right)
\!\!\left(U{\nGam}{^\alpha_{\sigma\mu}}U^{-\!1}\!
+\delta^\alpha_\sigma U_{,\mu}U^{-\!1}\!
-\delta^\alpha_\mu U_{,\sigma}U^{-\!1}\!\right)\nonumber\\
\fl &-&\!\!\frac{\lower2pt\hbox{$1$}}{(n\!-\!1)}
\left(U{\nGam}{^\tau_{[\tau\nu]}}U^{-\!1}\!+(n\!-\!1)\,U_{,\nu}U^{-\!1}\!\right)
\!\!\left(U{\nGam}{^\rho_{[\rho\mu]}}U^{-\!1}\!+(n\!-\!1)\,U_{,\mu}U^{-\!1}\!\right)\\
\fl &=&U\!\left(\nGam^\alpha_{\nu\mu,\alpha}
-\nGam^\alpha_{\!(\alpha(\nu),\mu)}
+\frac{\lower2pt\hbox{$1$}}{2}\nGam^\sigma_{\nu\mu}\nGam^\alpha_{\!(\sigma\alpha)}
+\frac{\lower2pt\hbox{$1$}}{2}\nGam^\alpha_{\!(\sigma\alpha)}\nGam^\sigma_{\nu\mu}
-\nGam^\sigma_{\nu\alpha}\nGam^\alpha_{\sigma\mu}
-\frac{{\nGam}{^\tau_{\![\tau\nu]}}{\nGam}{^\rho_{\![\rho\mu]}}}{(n\!-\!1)}\right)\!U^{-\!1}\nonumber\\
\fl &+&U_{,\alpha}{\nGam}{^\alpha_{\nu\mu}}U^{-\!1}
+U{\nGam}{^\alpha_{\nu\mu}}U^{-\!1}_{,\alpha}
+U_{,\mu}U^{-\!1}_{,\nu}-U_{,\nu}U^{-\!1}_{,\mu}\nonumber\\
\fl &-&\frac{\lower2pt\hbox{$1$}}{2}U_{,\mu}\nGam^\alpha_{\!(\alpha\nu)}U^{-\!1}
-\frac{\lower2pt\hbox{$1$}}{2}U\nGam^\alpha_{\!(\alpha\nu)}U^{-\!1}_{,\mu}
-\frac{\lower2pt\hbox{$1$}}{2}U_{,\nu}\nGam^\alpha_{\!(\alpha\mu)}U^{-\!1}
-\frac{\lower2pt\hbox{$1$}}{2}U\nGam^\alpha_{\!(\alpha\mu)}U^{-\!1}_{,\nu}\nonumber\\
\fl &+&\frac{\lower2pt\hbox{$1$}}{2}U_{,\mu}\nGam^\alpha_{\!(\nu\alpha)}U^{-\!1}
-\frac{\lower2pt\hbox{$1$}}{2}U_{,\nu}\nGam^\alpha_{\!(\mu\alpha)}U^{-\!1}\nonumber\\
\fl &-&\frac{\lower2pt\hbox{$1$}}{2}U\nGam^\alpha_{\!(\nu\alpha)}U^{-\!1}_{,\mu}
+\frac{\lower2pt\hbox{$1$}}{2}U\nGam^\alpha_{\!(\mu\alpha)}U^{-\!1}_{,\nu}\nonumber\\
\fl &+&U{\nGam}{^\sigma_{\nu\sigma}}U^{-\!1}_{,\mu}
\!-U{\nGam}{^\sigma_{\nu\mu}}U^{-\!1}_{,\sigma}
\!-U_{,\alpha}{\nGam}{^\alpha_{\nu\mu}}U^{-\!1}
\!+U_{,\nu}{\nGam}{^\alpha_{\alpha\mu}}U^{-\!1}
\!+(2\!-\!n)U_{,\nu}U^{-\!1}_{,\mu}\!-U_{,\mu}U^{-\!1}_{,\nu}\!\nonumber\\
\fl &+&\!U{\nGam}{^\tau_{[\tau\nu]}}U^{-\!1}_{,\mu}
-U_{,\nu}{\nGam}{^\rho_{[\rho\mu]}}U^{-\!1}
+(n\!-\!1)U_{,\nu}U^{-\!1}_{,\mu}\\
\label{hRtransformA}
\fl &=&U\hR_{\nu\mu}(\nGam)U^{-\!1}.
\end{eqnarray}
}

{
Substituting $\fl {\nGam}{^\alpha_{\nu\mu}}=\tGam{^\alpha_{\nu\mu}}+(\delta^\alpha_\mu\Am_\nu
\!-\delta^\alpha_\nu \Am_\mu)\,i\sqrt{\sixteenpid\Lambda_b}$ from (\ref{gamma_natural2}) into $\hR_{\nu\mu}$
and using $\tGam^\alpha_{\nu\alpha}\!=\!\nGam^\alpha_{\!(\nu\alpha)}\!=\!\tGam^\alpha_{\alpha\nu}$
from (\ref{JScontractionsymmetric2}) with the notation $[A,B]\!=\!AB\!-\!BA$ gives (\ref{breakout0}),
\begin{eqnarray}
\fl \hR_{\nu\mu}(\nGam)
&=&\nGam^\alpha_{\nu\mu,\alpha}-\nGam^\alpha_{\!(\alpha(\nu),\mu)}
+\frac{\lower2pt\hbox{$1$}}{2}\nGam^\sigma_{\nu\mu}\nGam^\alpha_{\!(\sigma\alpha)}
+\frac{\lower2pt\hbox{$1$}}{2}\nGam^\alpha_{\!(\sigma\alpha)}\nGam^\sigma_{\nu\mu}
-\nGam^\sigma_{\nu\alpha}\nGam^\alpha_{\sigma\mu}
-\frac{{\nGam}{^\tau_{\![\tau\nu]}}{\nGam}{^\rho_{\![\rho\mu]}}}{(n\!-\!1)}\\
\fl &=&\left(\tGam{^\alpha_{\nu\mu}}+(\delta^\alpha_\mu\Am_\nu
\!-\delta^\alpha_\nu \Am_\mu)\,i\sqrt{\sixteenpid\Lambda_b}\right)\!\!{_{,\alpha}}
-\tGam^\alpha_{\!(\alpha(\nu),\mu)}\nonumber\\
\fl &&+\frac{\lower2pt\hbox{$1$}}{2}\left(\tGam{^\sigma_{\nu\mu}}+(\delta^\sigma_\mu\Am_\nu
\!-\delta^\sigma_\nu \Am_\mu)\,i\sqrt{\sixteenpid\Lambda_b}\right)\tGam^\alpha_{\!(\sigma\alpha)}\nonumber\\
\fl &&+\frac{\lower2pt\hbox{$1$}}{2}\tGam^\alpha_{\!(\sigma\alpha)}\!\left(\tGam{^\sigma_{\nu\mu}}+(\delta^\sigma_\mu\Am_\nu
\!-\delta^\sigma_\nu \Am_\mu)\,i\sqrt{\sixteenpid\Lambda_b}\right)\nonumber\\
\fl &&-\left(\tGam{^\sigma_{\nu\alpha}}+(\delta^\sigma_\alpha\Am_\nu
\!-\delta^\sigma_\nu \Am_\alpha)\,i\sqrt{\sixteenpid\Lambda_b}\right)
\!\!\left(\tGam{^\alpha_{\sigma\mu}}+(\delta^\alpha_\mu\Am_\sigma
\!-\delta^\alpha_\sigma \Am_\mu)\,i\sqrt{\sixteenpid\Lambda_b}\right)\nonumber\\
\fl &&+\sixteenpid\Lambda_b(n\!-\!1)\Am_\nu\Am_\mu\\
\fl &=&\tGam^\alpha_{\nu\mu,\alpha}-\tGam^\alpha_{\alpha(\nu,\mu)}
+\frac{\lower2pt\hbox{$1$}}{2}\tGam^\sigma_{\nu\mu}\tGam^\alpha_{\sigma\alpha}
+\frac{\lower2pt\hbox{$1$}}{2}\tGam^\alpha_{\sigma\alpha}\tGam^\sigma_{\nu\mu}
-\tGam^\sigma_{\nu\alpha}\tGam^\alpha_{\sigma\mu}\nonumber\\
\fl &&+2\Am_{[\nu,\mu]}\,i\sqrt{\sixteenpid\Lambda_b}\nonumber\\
\fl &&+\frac{\lower2pt\hbox{$1$}}{2}(\delta^\sigma_\mu\Am_\nu
\!-\delta^\sigma_\nu \Am_\mu)\tGam^\alpha_{\sigma\alpha}i\sqrt{\sixteenpid\Lambda_b}\nonumber\\
\fl &&+\frac{\lower2pt\hbox{$1$}}{2}\tGam^\alpha_{\sigma\alpha}(\delta^\sigma_\mu\Am_\nu
\!-\delta^\sigma_\nu \Am_\mu)\,i\sqrt{\sixteenpid\Lambda_b}\nonumber\\
\fl &&-\tGam{^\sigma_{\nu\alpha}}(\delta^\alpha_\mu\Am_\sigma
\!-\delta^\alpha_\sigma \Am_\mu)\,i\sqrt{\sixteenpid\Lambda_b}\nonumber\\
\fl &&-(\delta^\sigma_\alpha\Am_\nu\!-\delta^\sigma_\nu \Am_\alpha)
\tGam{^\alpha_{\sigma\mu}}i\sqrt{\sixteenpid\Lambda_b}\nonumber\\
\fl &&+\sixteenpid\Lambda_b(n\!-\!1)\Am_\nu\Am_\mu
+\sixteenpid\Lambda_b((2\!-\!n)\Am_\nu\Am_\mu-\Am_\mu\Am_\nu)\\
\label{breakoutA}
&=&\hR_{\nu\mu}(\tGam)+2\Am_{[\nu,\mu]}\,i\sqrt{\sixteenpid\Lambda_b}
+\sixteenpid\Lambda_b[\Am_\nu,\Am_\mu]\nonumber\\
\fl &&+([\Am_\alpha,\tGam{^\alpha_{\nu\mu}}]
-[\Am_{(\nu},\tGam^\alpha_{\mu)\alpha}])\,i\sqrt{\sixteenpid\Lambda_b}\,.
\end{eqnarray}
}

\section{\label{ApproximateFandg}Approximate solution for $N_{\nu\mu}$ in terms
of $\sg_{\nu\mu}$ and $f_{\nu\mu}$}
Here we invert the definitions (\ref{sgdef2},\ref{fdef2}) of
$\sg_{\nu\mu}$ and $f_{\nu\mu}$ to obtain
(\ref{approximateNbar2},\ref{approximateNhat2}), the approximation of
$N_{\nu\mu}$ in terms of $\sg_{\nu\mu}$ and $f_{\nu\mu}$.
First let us define the notation
\begin{eqnarray}
\label{hfdef}
\hf^{\nu\mu}\!=\!f^{\nu\mu}i\sqrt{\sixteenpid}\,\Lambda_b^{\!-1/2}.
\end{eqnarray}
We assume that $|\hf^\nu{_\mu}|\!\ll\!1$ for all components of the
unitless field $\hf^\nu{_\mu}$, and find a solution
in the form of a power series expansion in $\hf^\nu{_\mu}$.

For the following calculations we will treat the fields as $nd\times nd$ matrices
but we will only show the tensor indices explicitly.
Lowering an index on the right side of the equation
$(\RmN/\Rmg)N^{\dashv\nu\mu}\!=\!\sg^{\mu\nu}\!+\!\hf^{\mu\nu}$ from (\ref{Wdef2}) we get
\begin{eqnarray}
\label{gminusF22}
(\RmN/\Rmg)N^{\dashv\mu}{_\alpha}
=\delta^\mu_\alpha I-\hf^\mu{_\alpha}.
\end{eqnarray}
Using $\hf^\alpha{_\alpha}\!=\!0$, the well known formula $det(e^M)=exp\,(tr(M))$,
and the power series $ln(1\!-\!x)=-x-x^2/2-x^3/3\dots$
we get\cite{Deif},
\begin{eqnarray}
\label{lndetspecial2}
ln(det(I\!-\!\hf))&=&tr(ln(I\!-\!\hf))
=-\frac{1}{2}\,tr(\hf^\rho{_\sigma}\hf^\sigma{_\rho})+(\hf^3)\dots
\end{eqnarray}
Here the notation $(\hf^3)$ refers to terms like
$tr(\hf^\tau{_\alpha}\hf^\alpha{_\sigma}\hf^\sigma{_\tau})$.
Taking $ln(det())$ on both sides of (\ref{gminusF22}) using the result (\ref{lndetspecial2}),
the definitions (\ref{RmgRmNdef}),
and the identities $det(sM^{})\!=s^{nd}det(M^{})$ and $det(M^{-1}_{})\!=1/det(M^{})$ gives
\begin{eqnarray}
\fl ln(det[(\RmN/\Rmg)N^{\dashv\mu}{_\alpha}])
=ln((N/\sg)^{n/2-1})
=-\frac{1}{2}\,tr(\hf^\rho{_\sigma}\hf^\sigma{_\rho})
+(\hf^3)\dots,\\
\fl ln[(\RmN/\Rmg)]
\label{lnapproxdetN2}
=-\frac{1}{2d(n\!-\!2)}\,tr(\hf^\rho{_\sigma}\hf^\sigma{_\rho})
+(\hf^3)\dots\,.
\end{eqnarray}
Taking $e^x$ on both sides of this and using $e^x=1+x+x^2/2\dots$ gives
\begin{eqnarray}
\label{approxdetN2}
(\RmN/\Rmg)
=1\!-\!\frac{1}{2d(n\!-\!2)}\,tr(\hf^\rho{_\sigma}\!\hf^\sigma{_\rho})+(\hf^3)\dots\,.
\end{eqnarray}
Using the power series $(1\!-\!x)^{-1}\!=\!1+x+x^2+x^3\dots$,
or multiplying by (\ref{gminusF22}) on the right
we can calculate the inverse of (\ref{gminusF22}) to get\cite{Deif}
\begin{eqnarray}
\fl (\Rmg/\RmN)N^\nu{_\mu}
=\delta^\nu_\mu I+\hf^\nu{_\mu}+\hf^\nu{_\sigma}\hf^\sigma{_\mu}+(\hf^3)\dots.
\end{eqnarray}
Lowering this on the left gives,
\begin{eqnarray}
\label{approxN2}
\fl N_{\nu\mu}
=(\RmN/\Rmg)(\sg_{\nu\mu}+\hf_{\nu\mu}+\hf_{\nu\sigma}\hf^\sigma{_\mu}+(\hf^3)\dots).
\end{eqnarray}
Here $(\hf^3)$ refers to terms like $\hf_{\nu\alpha}\hf^\alpha{_\sigma}\hf^\sigma{_\mu}$.
Using (\ref{cnumber},\ref{approxN2},\ref{approxdetN2},\ref{hfdef}) we get the result
(\ref{approximateNbar2},\ref{approximateNhat2}).

\end{document}